\documentclass[a4paper, aps, prb, amsmath, amssymb]{revtex4}
\usepackage[utf8]{inputenc}
\usepackage{graphicx, color}
\usepackage{multirow}
\usepackage{bm, soul}
\usepackage{algorithm, algorithmic}
\usepackage{subfigure}
\usepackage[raggedright]{sidecap}
\sidecaptionvpos{figure}{c} 

\newcommand{\Ipr}{\ensuremath{I_\mathrm{pr}}}
\newcommand{\Vpr}{\ensuremath{V_\mathrm{pr}}}
\newcommand{\Isat}{\ensuremath{I_\mathrm{sat}}}
\newcommand{\Vp}{\ensuremath{V_\mathrm{p}}}
\newcommand{\Vf}{\ensuremath{V_\mathrm{f}}}
\newcommand{\Te}{\ensuremath{T_\mathrm{e}}}
\newcommand{\nee}{\ensuremath{n_\mathrm{e}}}
\newcommand{\neng}{\ensuremath{\overline{n}_\mathrm{e}/n_\mathrm{G}}}

\newcommand{\mus}{\ensuremath{\mu \mathrm{s}}}
\newcommand{\sigte}{\ensuremath{\sigma_{\Te} / \Te}}
\newcommand{\deltaV}{\ensuremath{\triangle_V}}
\newcommand{\Tetwoms}{\ensuremath{\overline{T}_{\mathrm{e}, 2 \mathrm{ms}}}}

\newcommand{\Eqnref}[1]{Eq.(~\ref{eq:#1})}
\newcommand{\tabref}[1]{tab.~\ref{tab:#1}}
\newcommand{\Figref}[1]{Fig.~\ref{fig:#1}}

\newcommand{\secref}[1]{Sec.~\ref{sec:#1}}

\newcommand{\argmin}{\operatornamewithlimits{arg\ min}}

\begin{document}
\title{Outlier classification using Autoencoders: application for fluctuation driven flows in fusion plasmas}
\author{R.~Kube}
\email{ralph.kube@uit.no}
\author{F.~M.~Bianchi}
\affiliation{Department of Physics and Technology, UiT The Arctic University of Norway, N-9037 Tromsø, Norway}
\author{D.~Brunner}
\affiliation{Commonwealth Fusion Systems, Cambridge, MA, USA}
\author{B.~LaBombard}
\affiliation{MIT Plasma Science and Fusion Center, Cambridge, MA, 02139, USA}

\date{\today}
\begin{abstract}
    Understanding the statistics of fluctuation driven flows in the boundary layer of
    magnetically confined plasmas is desired to accurately model the lifetime of the
    vacuum vessel components. Mirror Langmuir probes (MLPs) are a novel diagnostic that
    uniquely allow to sample the plasma parameters on a time scale shorter than the
    characteristic time scale of their fluctuations. Sudden large-amplitude fluctuations
    in the plasma degrade the precision and accuracy of the plasma parameters reported by
    MLPs for cases in which the probe bias range is of insufficient amplitude. While some data
    samples can readily be classified as valid and invalid, we find that such a
    classification may be ambiguous for up to $40\%$ of data sampled for the plasma parameters
    and bias voltages considered in this study.
    In  this contribution we employ an autoencoder (AE) to learn a low-dimensional representation of
    valid data samples. By definition, the coordinates in this space are the features that mostly
    characterize valid data.
    Ambiguous data samples are classified in this space using standard classifiers for vectorial 
    data. This way, we avoid to define complicate threshold rules to identify outliers, 
    which requires strong assumptions and introduce biases in the analysis. Instead, these rules
    are learned from the data by statistical inference.
    By removing the outliers that are identified in the latent low-dimensional space of the AE,
    we find that the average conductive and convective radial heat flux are between approximately 
    $5$ and $15\%$ lower
    as when removing outliers identified by threshold values. 
    For contributions to the radial heat flux due to triple correlations, the difference is up to
    $40\%$.  
\end{abstract}

\maketitle


\section{Introduction}
Tokamaks confine fusion plasmas, a fully ionized hydrogen plasma with a core
temperature of approximately $100,000,000\, K$, using strong, donut-shaped magnetic
fields within a vacuum vessel \cite{tokamaks}. The outer boundary region of the plasma
comprises a region where closed magnetic field lines wind around toroidal surfaces
and a region where open magnetic field lines are guided as to intersect material walls, 
so-called divertor targets, remote from the central plasma column. As plasma streams along
the open field lines onto the divertor targets, it cools. These field lines terminate at divertor
structures which facilitate the further removal of the plasma. Thereby this region defines an
exhaust channel for the plasma. Intermittent, large-amplitude fluctuations of the plasma parameters,
such as the density and the temperature, are characteristic for the outboard mid-plane open
field line region \cite{antar-2001prl, antar-2003, xu-2005, graves-2005, garcia-2013}.
These fluctuations are foot prints of coherent structures of excess plasma pressure, called blobs,
which propagate radially out over through the open field line region onto the vacuum vessel walls 
at the outboard mid-plane \cite{terry-2003, zweben-2004, terry-2005, agostini-2011, kube-2013}.
Depending on their amplitude, these fluctuations can potentially erode the vacuum vessel. Impurities
released from the wall may furthermore accumulate within the confined plasma column and negatively
impact the confinement properties of the plasma. Todays tokamaks perform experiments on plasma
discharges which last for several seconds. Future fusion reactors need to operate with long
pulses or continuously. In order to model the life time of the plasma facing components for
such requirements, a precise and accurate description of this fluctuation driven transport is
desired \cite{federici-2001, whyte-2009}.

Langmuir probes are the workhorse used to diagnose this boundary region plasma. They are 
implemented as electrodes immersed into a plasma. Using electric current and voltage samples
recorded by a Langmuir probe, plasma quantities are recovered from the relation \cite{hutch-book}
\begin{align}
    \Ipr = \Isat \left[ 1 - \exp \left(\frac{\Vpr - \Vf}{\Te} \right)\right]. \label{eq:Ipr}
\end{align}
Here $\Ipr$ is the collected electric current and $\Vpr$ applied bias voltage. $\Te$ gives the electron
temperature of the plasma. The floating potential $\Vf$ is defined as the electric potential assumed
by an electrically isolated object were it to be immersed into the sampled plasma. The ion saturation
current $\Isat$ is the maximal current that can be drawn by an electrode, which is limited by ion
collection of the electrode.

In order to estimate the particle and heat fluxes driven by the electric drift,
the electron density, temperature, and the local electric field need to be recovered
from probe measurements. Commonly, these quantities are recovered from probes by applying
a sweeping voltage to the electrode. This allows to sample several current-voltage
measurements  $\left( \Ipr, \Vpr \right)$ during one sweep. From these, 
$\Isat$, $\Te$ and $\Vf$ are obtained from a fit on \Eqnref{Ipr}.
The ion saturation current and the electron temperature can be used to calculate the
electron density of the plasma as \cite{hutch-book}
\begin{align}
    \nee = 2 \frac{\Isat}{e A_\mathrm{p} \sqrt{k_\mathrm{b} \Te / m_\mathrm{i}}}. \label{eq:ne}
\end{align}
Here $e$ is the elementary charge, $A_\mathrm{p}$ is the current collecting area of
the electrode, $k_\mathrm{b}$ is the Boltzman constant, and $m_\mathrm{i}$ denotes the
ion mass. The electric potential in the plasma can be estimated as 
\begin{align}
    \Vp = \Vf + \Lambda \Te,
\end{align}
where $\Lambda \approx 2-3$ for scrape-off layer plasmas \cite{stangeby-book, rohde-1996}. 
Potential measurements from poloidally separated electrodes allow to estimate
the poloidal electric field, which drives the radial electric drift.

A characteristic time scale for fluctuations of $\nee$, $\Te$, and $\Vp$ in
boundary plasma is given by approximately $10 \mus$
\cite{boedo-2001, boedo-2003, kirnev-2004, garcia-2006-tcv, horacek-2010, garcia-2013,
garcia-2016-nme, kube-2016-ppcf, theodorsen-2016-ppcf}. 
Sweeping the voltage with a frequency larger than approximately $100\, \mathrm{kHz}$ however
leads to hysteresis effects in the sampled current-voltage
characteristic as the bias voltage polarizes the flux tube that plasma is sampled from 
\cite{muller-2010, verplancke-1996}. Thus, Langmuir Probes used in this manner can not
sample the plasma parameters on a fast enough time scale to resolve the fluctuations of
the boundary layer plasma.

The Mirror Langmuir probe (MLP) biasing technique allows to sample $\Isat$, $\Te$, and $\Vf$
on a time scale below that of the boundary layer plasma fluctuation
\cite{labombard-2007, labombard-2014}. The MLP diagnostic consists of three main components.
The actual mirror Langmuir probe is an electric circuit outputs a current-voltage (I-V)
characteristic with three adjustable parameters $\Isat$, $\Te$, and $\Vf$, given by \Eqnref{Ipr}.
The second main component is a Langmuir electrode immersed in the plasma. Both components are connected
to a fast switching biasing waveform, the third main component of the MLP diagnostic. The bias
waveform switches between the states $(V^{+}, V^{0}, V^{-})$, such that the Langmuir electrode
draws approximately $\pm \Isat$ at the states $V^{\pm}$ and zero net current when biased to
$V^{0}$, as shown in Fig. 1 of \cite{labombard-2014}. The target bias voltage state is
updated every $300\, \mathrm{ns}$. Once the bias voltage has settled, the current drawn from the
MLP and the Langmuir electrode are sampled. The ion saturation current, the plasma potential,
and the electron temperature are recovered by a fit of \Eqnref{Ipr} to the data samples from the 
Langmuir electrode. 

The main task of the MLP circuit is to set and maintain the optimal range
of the bias voltages such that a complete $I-V$ characteristic can be reconstructed from
current samples drawn by the Langmuir electrode at the three bias voltage states.
In order to account for varying plasma conditions, the MLP dynamically updates the voltage states
$V^{+}$ and $V^{-}$ relative to the running average of the electron temperature samples over a
$2\, \mathrm{ms}$ window such that $\deltaV < 4 \Tetwoms$ holds. Here, $\deltaV = V^{+} - V^{-}$
and $\Tetwoms$ denotes this running average of the electron temperature. 

Large amplitude fluctuations of the boundary layer plasma on the other hand have a characteristic
time scale of approximately $10 \mus$. During such transient events, the electron temperature may
significantly exceed the running average, $\Te > \Tetwoms$ such that the adjusted biasing voltage
range may be insufficient to guarantee a precise fit on the true $I-V$ characteristic of a
hypothetical Langmuir probe. But also events such as probe arcing may result in unphysical fit
values.

A large body of experimental measurements suggest that the fluctuation statistics of the
boundary plasma depend on the global parameters of the plasma discharge, such as line-averaged
core plasma density and the magnetic geometry 
\cite{labombard-2001, greenwald-2002, boedo-2003, garcia-2007-nf, carralero-2014, kube-2016-ppcf}.
Since the MLP biasing drive is agnostic to these circumstances, the accuracy and precision of 
data samples reported by the MLP may vary, depending on the plasma it samples. In order to 
accurately calculate lower order statistical moments of MLP data time series or distributions
such as the probability distribution function or power spectral density, low-accuracy data samples
should be discarded. However, if too many samples are discarded, these moments or distribution
functions cannot be estimated with high statistical significance, due to the scarceness of
available data points.

One way of pruning MLP data time series is to define valid ranges for the MLP parameters.
Within these thresholds, samples are kept and out of bounds samples are to be discarded.
A sensible boundary, or thresholds, needs to be low enough in order to reject samples with
unphysically large fluctuation values. On the other hand, the threshold value needs to be
large enough so that the accepted data points correctly capture the properties of the plasma
fluctuations of the interrogated plasma. 
While measurements with a sufficient or insufficient biasing voltage range are
readily identifiable, such a decision is ambiguous for a large fraction of other samples.
In practice, it is often the case that several nearby Langmuir electrodes sample the plasma.
Given that MLP samples may be quiet heterogeneous when operating on a small biasing voltage
range, a threshold based method requires domain expertise and inevitably introduces biases.

\subsection*{Proposed Approach}

The approach proposed here adopts simple thresholds to identify all \emph{good} and \emph{bad}
measurements as a primer. This identification will be non-exhaustive, and will leave a large
number of samples unclassified. From this, all uncertainty in the quality of the measurements will
be treated with machine learning techniques which exploit statistical properties and regularities 
in the data. This approach allows to label unclassified data by making inference, as opposed to
labing them using a complicated set of rules.

Specifically, we present an outlier classification framework based on an autoencoder (AE), a type of
neural network that can be used to learn low-dimensional representations of arbitrary datasets.
AEs will be trained using only good measurements samples so that they learn how to map them into low
dimensional representations. Each dimension of the space induced by the AE mapping corresponds to a
combination of features which best characterize the important features of \emph{good} measurements.
Those features are identified without making any \textit{a-priori} assumption, but are automatically
selected by the AE as the ones that are, \textit{on average}, the most informative to describe the
training samples. As a consequence, the numerical values of features in training samples will
be similar and are mapped into a compact cluster in that low dimensional space.

AEs learn a representation of \emph{good} measurement that are more \textit{powerful}, due to the
regularization constraints of the dimensionality reduction, and generalize better the samples.
Evaluating similarities among samples represented in this new space is arguably more meaningful and
reliable.

Once an AE is trained and the mapping to such a low dimensional space is learned, the unclassified
samples will be processed. \emph{Bad} measurements lack the characteristic features of \emph{good}
measurements and are expected to map onto vectors with a large distance to the cluster composed of
\emph{good} samples.

In order to identify a boundary between the representations of \emph{good} and \emph{bad}
measurements, classifiers for vectorial data will be trained in this new space. Unclassified data
samples are assigned a label based on which side of the decision boundary they fall.

The rest of this article is structured as follows: \secref{measurements}
describes measurements of plasma fluctuations by MLPs and discusses the structure of
valid and invalid data at hand. \secref{autoencoder} introduces AEs and desribes their application for
outlier detection in large datasets. The proposed classification method and its application to
MLP data is described in \secref{classification}. \secref{performance} discusses the performance
of the proposed framework and \secref{conclusion} gives a conclusion.


\section{Measurements of plasma fluctuations}
\label{sec:measurements}
\begin{figure}
    \centering
    \begin{minipage}{0.45\textwidth}
    \includegraphics[width=\textwidth]{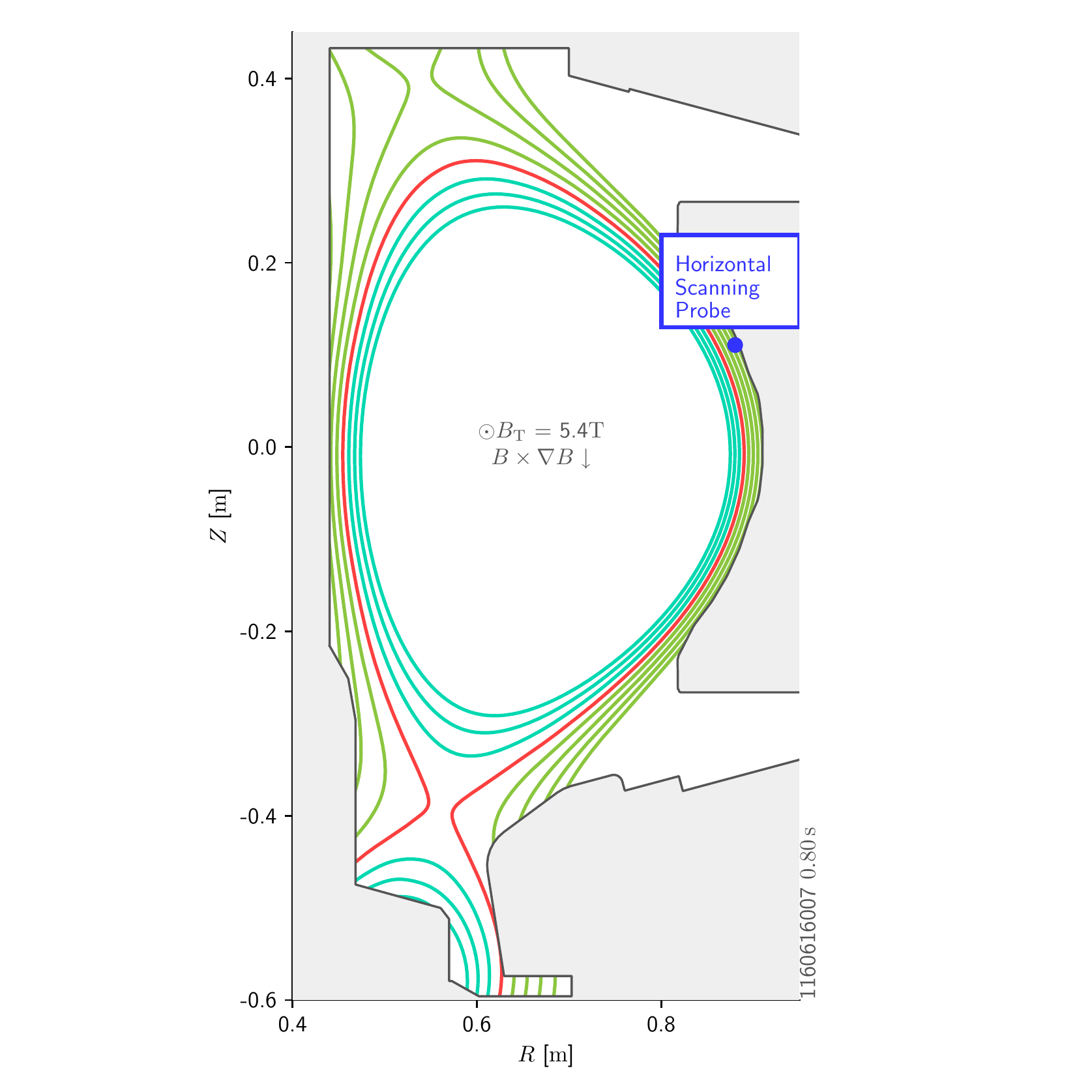}
    \caption{The poloidal cross-section of the Alcator C-Mod tokamak. The blue dot
             marks the location where the MLPs sample the plasma. Green lines denote the
             open magnetic field lines, cyan lines denote the closed magnetic field lines.
             The red line separates the open field line region from the closed field line region.
             Material structures are shown in gray.}
    \label{fig:cmod_cx}
    \end{minipage}
    \begin{minipage}{0.45\textwidth}
        \includegraphics[width=0.5\textwidth]{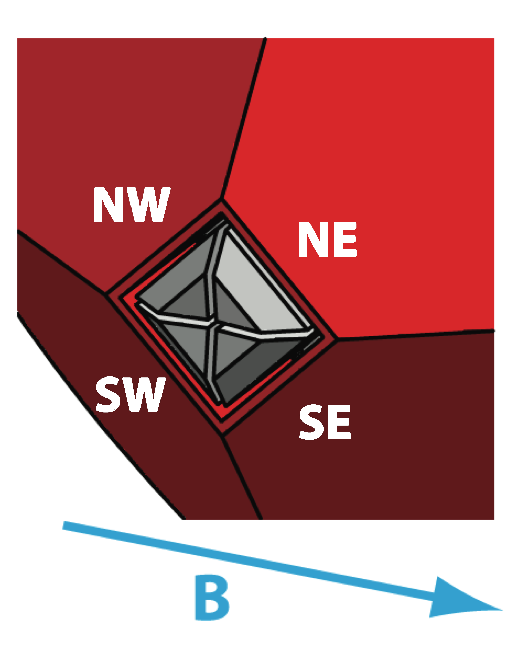}
        \caption{The Mach probe head with four Langmuir electrodes, labelled ''NE'', ''SE'',
                 ''SW'', and ''SE'', protruding from its top. The blue arrow denotes the direction
                 of the local magnetic field.}
        \label{fig:probe_head}
    \end{minipage}
\end{figure}

Dedicated experiments with the goal to describe the statistics of fluctuation driven flows in the
boundary plasma have been performed in the Alcator C-Mod tokamak 
\cite{hutchinson-1994, greenwald-2013, greenwald-2014, kube-2018-ppcf}. 
In these experiments, the boundary layer of an ohmically heated, lower single-null plasma discharge 
with a toroidal magnetic field strength of $B_\mathrm{T} = 5.4\, \mathrm{T}$, was interrogated by
four MLPs, connected to the electrodes of a Mach probe head, as shown in \Figref{probe_head}.
The probe head was mounted on a linear servomotor probe drive system \cite{brunner-2017-rsi} and
dwelled flush with plasma facing components at the outboard mid-plane location,
as shown in \Figref{cmod_cx}. Extraordinarily long data time series of one second duration
were sampled in stationary plasma discharge conditions with the goal to calculate the fluctuation
statistics for the plasma with unprecedented accuracy. 

The line-averaged core plasma density of the investigated discharge is 
$\neng \simeq 0.6$, where $n_\mathrm{G}$ denotes the Greenwald
density \cite{greenwald-2002}. For such high line-averaged core plasma densities, the average
electron temperature in the far scrape-off layer plasma is below $10\, \mathrm{eV}$. For
lower $\neng$, the scrape-off layer is commonly warmer \cite{labombard-2001}.
On the other hand, the MLPs registers order unity fluctuations of the electron temperature. 
That is, for such high $\neng$ and accompanying temperatures in the scrape-off layer, 
the MLP biasing drive operates at the limits of its design.

In order to assess the accuracy of fit parameters reported by the MLPs, they were
compared among the four MLPs. Since the electrodes on the probe head are separated by
approximately $2\, \mathrm{mm}$, smaller than the characteristic size of structures
in the boundary layer \cite{kube-2013}, it is expected that all four MLPs report 
similar fit parameters.
Indeed, $\Isat$, $\Te$, and $\Vf$ fit parameters reported from the four MLPs are of 
comparable magnitude when the range of the biasing voltage states are large,
$\deltaV > 4 \Te$. For the case where $\deltaV \lesssim 4 \Te$,
the reported $\Te$ fit parameters may show significant deviations. 
Operating with small bias voltage ranges, the relative fit error of the electron temperature,
$\sigte$, is furthermore on average larger than for the case $\deltaV > 4 \Te$.
The relative error on $\Isat$ and $\Te$ reported by the fit routine are correlated with
a Pearson sample correlation coefficient of approximately one. The relative error on the
floating potential is uncorrelated to the relative error of both, $\Isat$ and $\Te$. While
both $\Isat$ and $\Te$ are positive definite quantities, $\Vf$ may assume both positive
as well as negative values. Thus, the relative error on the floating potential,
$\sigma_{\Vf} / \Vf$, assumes large absolute values for small absolute values of $\Vf$. 
This quantity is therefore not suitable to identify poor fits. Poor fits are identified by 
a large $\Te$ value, a large relative fit error $\sigma_{\Te} / \Te$, and a small fit domain
$\deltaV / \Te$. 

\begin{figure}[h!tb]
    \includegraphics{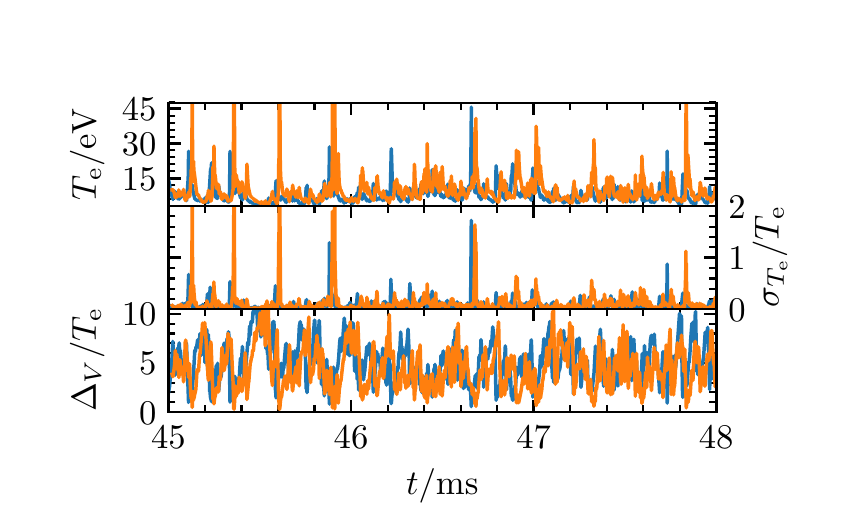}
    \caption{Time series of the electron temperature (upper panel), 
             the relative error on $\Te$ (middle panel), 
             and the range of the biasing voltages (lower panel),
             reported by the north-east (blue lines) and south-west (orange lines) MLP.
             Time series from the latter MLP are delayed by $20\, \mus$ for better
             visibility.}
    \label{fig:fit_timeseries}
\end{figure}

Figure \ref{fig:fit_timeseries} shows data time series reported by the north-east and
south-west MLP. The upper panel shows the electron temperature, the middle panel shows
the relative error on $\Te$, and the lower panel shows the biasing voltage range. 
A large fraction of the samples feature small to moderate $\Te$ values, together with small
error proxies, that is a relative error $\sigte \lesssim 0.1$ and large biasing voltage range.
Within these ranges, the fit parameters reported by the different MLPs are similar to one
another, indicating that they are both, accurate and precise.

Large-amplitude fluctuations of the electron temperature appear intermittently
in both time series. While the MLPs register them simultaneously, they report dissimilar
$\Te$ values, varying by up to $100\%$. Large amplitude fluctuations are furthermore
associated with a large relative error $\sigte$ and a small biasing voltage range. Comparing
the appearance of large amplitude peaks sampled by the two MLPs, they may be grouped into
several categories. One category are large amplitude peaks recorded by multiple MLPs but
with disparate $\Te$ values, for example at $45.1\, \mathrm{ms}$, at $45.4 \, \mathrm{ms}$,
or at $45.9\, \mathrm{ms}$. Another category are peaks where the MLPs report similar $\Te$
values, for example at $45.25\, \mathrm{ms}$ or at $46.6\, \mathrm{ms}$. Judging by the fit
parameters reported by a single MLP, such peaks should be discarded. However, in the case
where multiple MLPs report similar peaks, such samples may be retained. For the data at hand,
electrode-averaged values may be used in combination with threshold values to identify samples
which should be certainly kept or discarded. But for a majority of the data, such a simple
classification may be ambiguous. 

This broad range of variations under which large amplitude peaks are observed suggest that
it is impractical to develop a comprehensive set of rules based on wich to accept or reject 
reported peak amplitudes. In the following, we discuss how statistical inference can be used
to derive such rules based from a priming sample of \emph{good}, or accepted data.

\subsection*{Dataset description and threshold definition}

Data time series of $\Te$, $\sigte$, and $\deltaV$, sampled by all four MLPs, are combined into a single dataset
$\mathcal{X} = \left\{ T_{\mathrm{e},p},
                      \sigma_{T_{\mathrm{e}, p}} / T_{\mathrm{e}, p},
                      \triangle_{V} / T_{\mathrm{e}, p} \mid  p \in \left\{ \mathrm{NE}, \mathrm{SE}, \mathrm{SW}, \mathrm{NW} \right\} \right\}$.
Each sample is a vector in $\mathbb{R}^{12}$ corresponding to the individual measurements at a given
time.
We apply a simple threshold mechanism to label only a fraction of the original dataset. In
particular, we identify \emph{good} and \emph{bad} samples, $\mathcal{X}^g$ and $\mathcal{X}^b$, while
the remaining samples are left unlabelled and referred as \textit{uncertain} $\mathcal{X}^u$.

A fit reported by a single MLP is considered valid if $\Te$ and $\sigte$ are below a threshold
value, and if $\deltaV / \Te$ exceeds a 
threshold value. If the opposite conditions are true, the fit is considered invalid. 
If at least two MLPs report a valid fit, the vector is labelled \textit{good} and assigned 
to $\mathcal{X}^g$. 
If at least two MLPs report an invalid fit, the vector is labelled \textit{bad} and assigned 
to $\mathcal{X}^b$.

Table \ref{tab:good_bad_thresholds} lists three different sets of threshold values that
are used for an a-priori partitioning of the data set $\mathcal{X}$. Depending on the
threshold values used, the fraction of data points classified as \emph{good},
\emph{uncertain}, and \emph{bad} varies.
For example, the category \textit{relaxed} denotes the partitioning that excludes the least
amount of data from being categorized. Fits that report electron temperatures of up to 
$45\, \mathrm{eV}$
with a relative error of $0.75$ over a range of $\deltaV / \Te \geq 2.5$ are considered as valid.
The fraction of \emph{bad} and \emph{uncertain} samples are listed in bottom row of \tabref{good_bad_thresholds}.
Using \textit{relaxed} thresholds, approximately $20\%$ of the data is unclassified, while approximately
$40\%$ of the data is labeled \emph{uncertain} when using \textit{strict} thresholds. 

\begin{table}[h!tb]
    \begin{tabular}{c|c|c|c}
        Quantity                & relaxed               & mid                   & strict \\ \hline
        $\Te / \mathrm{eV}$     & $45/50$               & $40/45$               & $35/40$       \\ 
        $\sigte$                & $0.75/1.0$            & $0.5/0.75$            & $0.25/0.5$    \\
        $\deltaV / \Te$         & $2.5/1.5$             & $3.0/2.0$             & $3.5/2.5$     \\ \hline
        \emph{uncertain}/\emph{bad}   & $20.3\%$ / $0.1\%$    & $30.0\%$ / $0.1\%$    & $40.2\%$ / $0.2\%$
    \end{tabular}
    \caption{Threshold values used for a-priori partitioning of the data.
             The first number gives the threshold for a poor fit, the second number gives 
             the threshold for a good fit. The lowest row lists the fraction of data labeled as
             \emph{uncertain} and $\emph{bad}$.}
    \label{tab:good_bad_thresholds}
\end{table}

In the following, we describe an approach where an AE is facilitated to identify data,
which cannot be classified reliably by applying a threshold method.

\section{Autoencoders}
\label{sec:autoencoder}

AEs~\cite{Hinton504} are a particular class of neural networks, which received
increasing interest in recent years~\cite{kingma2013auto,makhzani2015adversarial,bianchi2017learning}.
AEs can be used to learn unsupervised compressed, or lossy, representations of data, by training the
network to map the input in a lower dimensional space through a bottleneck layer and then reconstruct
the original input. In this way, the AE learns how to compress inputs, by retaining only the most important
information necessary to yield a reconstruction that is as much accurate as possible~\cite{bengio2009learning}. 
Indeed, training AEs by minimizing
a reconstruction error corresponds to maximizing the lower bound of the mutual information between input
and the learned representation~\cite{vincent2010stacked}.

The bottleneck enforces a strong regularization that provides noise filtering, prevents the AE from learning
trivial identity mappings (i.~e., the identity function), and guarantees robustness to small changes in the
inputs~\cite{6472238}. Further regularization can be used to prevent overfitting on the training data and
enhance the generalization properties of the representations. The most common regularizations are applying 
a $\ell_2$ norm penalty to the weights learned network, and using dropout~\cite{srivastava2014dropout} to 
randomly drop connections between neurons at each iteration in the training phase. 
Dropout hinders couplings among neurons and therefore encourages to diversify the behavior of neurons.

\begin{SCfigure}[1.5][th!]
    \centering
    \includegraphics[keepaspectratio,width=0.3\textwidth]{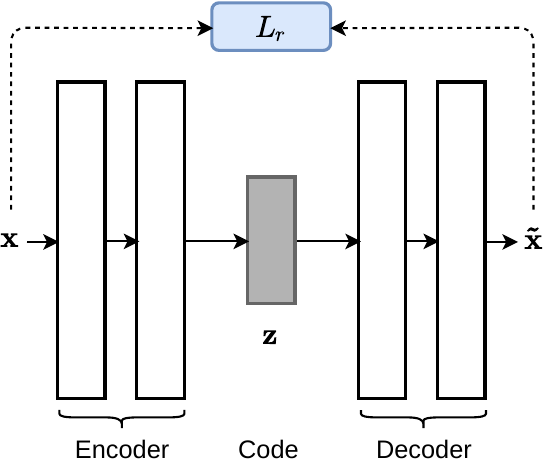}	
    \caption{Schematic representation of the AE architecture with a bottleneck. The encoder generates a low dimensional representation $\mathbf{z}$ of the input $\mathbf{x}$. The AE is trained by minimizing the discrepancy (quantified by the loss $L_r$) between $\mathbf{x}$ and its reconstruction $\mathbf{\tilde{x}}$ yielded by the decoder.}
    \label{fig:ae_arch}
\end{SCfigure}

In the training phase, an AE learns two functions at the same time. The first one is called \textit{encoder}
and provides a mapping from an input domain, $\mathcal{X}$, to a code domain, $\mathcal{Z}$, i.~e.~ the latent
representation space. Specifically, an input $\mathbf{x}$ is represented as the output $\mathbf{z}$ of the
innermost layer in the AE. The second function, called \textit{decoder}, implements a mapping from
$\mathcal{Z}$ back to $\mathcal{X}$. Fig.~\ref{fig:ae_arch} depicts a standard AE architecture with a bottleneck.

The encoding function $E(\cdot): \mathcal{X} \rightarrow \mathcal{Z}$ and the decoding function $D(\cdot): \mathcal{Z} \rightarrow \mathcal{X}$ of the AE define the following deterministic posteriors
\begin{align}
    \begin{split}
    \label{eq:encoding_decoding}
    \mathbf{z} &= E(\mathbf{x}) = p(\mathbf{z}|\mathbf{x};\theta_E) \\
    \mathbf{\tilde{x}} &= D(\mathbf{z}) = q(\mathbf{\tilde{x}}|\mathbf{z};\theta_D),
    \end{split}
\end{align}
where $\theta_E$ and $\theta_D$ are the trainable parameters of the two functions; $\mathbf{x}$ is the original
input; $\mathbf{z}$ is the code representation; $\mathbf{\tilde{x}}$ is the reconstruction of the input. The
encoding and decoding function are usually implemented as two feed-forward neural networks, which are constrained
to be symmetric. Each network consists of a stack of layers that can be dense, convolutional~\cite{10.1007/978-3-642-21735-7_7} or recurrent~\cite{bianchi2017recurrent}.
Here, we focus only on dense layers that are implemented by an affine transformation
followed by a non-linear activation function applied component-wise. Common activation functions are the sigmoid
(logistic function, \textit{tanh}), the maxout ~\cite{goodfellow-2013}, and the rectified linear unit (ReLU).

Each layer contains a different number of processing units (neurons), which affects the capability of approximating
a generic function. While a large number of layers and neurons per layer can provide more powerful modeling capabilities,
the number of parameters increases with a consequent risk of overfit and a greater demand of computational resources.
Therefore, an optimal configuration of the network should account for those contrasting properties.

The configuration of an AE with $K$ layers in the encoder and decoder, respectively, can be suitably expressed as
\begin{equation}
    \label{ae:config}
    \mathcal{C} = \{ e_0, \dots, e_K, z, d_0, \dots, d_K \},
\end{equation}
where $e_i$ and $d_i$ define the number of neurons in the $i$-th layer of the encoder and the decoder. The size of the
innermost layer is denoted by $z$ and defines the dimension of the representation $\mathbf{z}$. As previously stated, we
implement a symmetric encoder/decoder architecture by enforcing the following constraint $e_i = d_{K-i}$.

In order to minimize the discrepancy between the input and its reconstruction, the parameters $\theta_E$ and $\theta_D$
are adjusted by minimizing through stochastic gradient descent the following reconstruction loss
\begin{equation}
    \label{eq:distortion}
    L = L_r + \lambda L_2 = \mathbb{E}_{\mathbf{x} \sim \mathcal{X}} \left[ \lVert \mathbf{x} - \mathbf{\tilde{x}} \rVert^{2} \right] + \lambda \left( \lVert \theta_E \rVert^2 + \lVert \theta_D \rVert^2 \right)\; .
\end{equation}
The term $L_r$ minimizes the mean squared error between original inputs and their reconstructions, while $L_2$ penalizes
large model weights. The hyperparameter $\lambda$ controls the latter contribution to the total loss.

Besides the regularization parameter $\lambda$ and the network configuration $\mathcal{C}$, other hyper-parameters that must
be chosen by the user, or optimized by means of a validation procedure, are the following: 
the probability $p_\text{drop}$ to drop neural connections during the training; 
the learning rate $\eta$ used in stochastic gradient descent; 
the type of activation function implementing the non-linearities within each layer of the AE. 
We refer to the whole set of hyper-parameters as $\Gamma_\text{ae}$.

\subsection*{Outlier detection with Autoencoders}

Outlier detection (also referred to as anomaly detection) is an important area of study in machine learning 
and is applied to several case-studies where non-nominal samples are scarce, noisy and not always available
during training. The objective of outlier detection procedures is to identify anomalous patterns, the outliers,
in data that do not conform to an expected behavior~\cite{chandola2009anomaly}.

Dimensionality reduction procedures, such as Principal Component Analysis (PCA), AEs and energy based
models~\cite{pmlr-v48-zhai16, sakurada2014anomaly} identify a subspace defined by the directions with largest
variation among the nominal samples. 
While PCA can only capture variations that emerge from linear relationships in the data, more sophisticated
models such as AEs account also for non-linear relationships. Therefore, AEs can identify a subspace defined
by features that better characterize the nominal samples.

Anomaly detection methods based on dimensionality reduction rely on the assumption that anomalous samples do
not belong to the subspace, learned during training, that contains nominal data. Indeed, the representations
generated for samples of a new, unseen class will arguably fail to retain important characteristics, since the
latent low-dimensional space induced by the AE does not span the most relevant features of the anomalous data. 
As direct consequences, the AE would yield large reconstruction errors for those samples and their
low-dimensional representations would be significantly different and more scattered than for samples from
the nominal class. This effect can be exploited to obtain an implicit separation between the classes
in the code space, which can facilitate the separations of the two classes by a subsequent classifier.

Similar assumptions are reasonable for the MLP dataset at hand. As shown in \Figref{fit_timeseries},
a large fraction of the MLP samples feature similar $\Te$ fit values, together with $\sigte$ and $\deltaV$
values which indicate a reliable fit. These samples are considered as inliers and are used to train an AE.
Having learned the important characteristics of inlier samples, hitherto unclassified samples
will be mapped into the code space of the AE. Samples which do not share the important characteristics
of the inlier samples should then be readily identifiable.
In the following, we describe a classification framework that exploits this property of the data at hand
to identify and separate outliers.

\section{Proposed classification framework and selection of model parameters}
\label{sec:classification}

The critical components of the proposed classification framework are the AE and the classifier used in the latent
code space of the AE to discriminate between \emph{good} and \emph{bad} samples.
Beside the trainable parameters, both components depend on a set of hyper-parameters whose tuning may affect the
behavior of the whole framework. In the following, we discuss how the choice of a classifier and hyper-parameters
for both, the AE and the classifier, results in different statistics of the inlier $\Te$ data. 
In the last section, we discuss how the choice results in different statistics of the
fluctuation driven heat flux. Since there is no
ground truth available, that is, the real electron temperature of the plasma is unknown, no quantitative evaluation
of the classification frameworks performance can be formulated. Instead, the design choices will be guided by the
inferred biases of the filtered datasets for any given set of hyperparameters of the classification framework.

\begin{figure}[th!]
    \centering
    \includegraphics[keepaspectratio,width=0.6\textwidth]{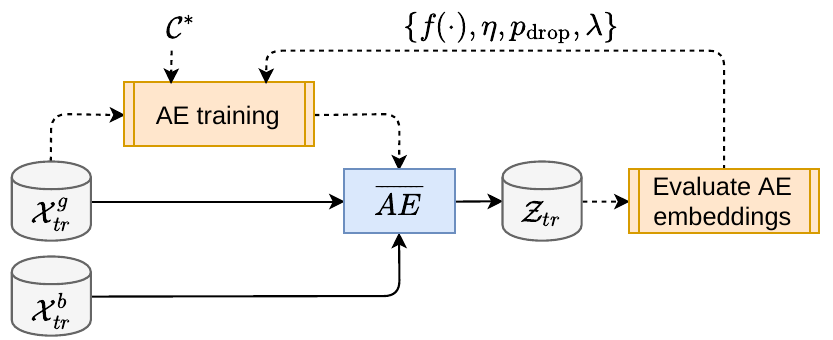}	
    \caption{Pipeline of the procedure to train the AE and select the optimal hyperparameters. 
             $\overline{AE}$ denotes a trained AE. $\mathcal{C}^*$ denotes an optimal layout of the AE that, 
             that is assumed to be given for the task to identify the optimal hyperparameters.}
    \label{fig:ae_train}
\end{figure}

As discussed in \secref{autoencoder}, the AE depends on several hyperparameters $\boldsymbol{\Gamma}_\mathrm{AE}$. 
In the following, we discuss the sensitivity of the mapping induced by the AE on them. Figure \ref{fig:ae_train} shows
the pipeline used for this task. For the sake of simplicity, we furthermore only consider the network layout 
$\mathcal{C}^{*} = \{12, 2, 12\}$ at this point.

$5000$ random elements from $\mathcal{X}^{g}$ are used to train the AE.  During training we 
observe little sensitivity to the hyper-parameters $p_\text{drop}$, $\eta$, and $\lambda$.
In the following, we select  $p_\mathrm{drop} = 10^{-1}$, $\eta = 10^{-2}$ and $\lambda = 10^{-3}$
as hyperparameters.

In feed-forward neural networks, each neuron computes the sum $y = \sum_{i} x_i w_i + b$,
where $x_i$ denotes the input from the previous layer, $w_i$ denotes a weight and $b$ denotes a
bias. The weights and the bias are determined during the training phase. The output of each
neuron is $f(y)$, which is called the activation function. The activation functions considered here are
%
\begin{equation}
\label{eq:activ_fun1}
    \begin{aligned}
    \text{sigmoid:} \;\; f(y) &= \frac{1}{1+e^{-y}}, \\
    \text{tanh:} \;\; f(y) &= \frac{1 - e^{-2y}}{1 + e^{-2y}}, \\
    \text{ReLU:} \;\; f(y) &= \max(0, y), \\
    \text{Maxout:} \;\; f(y) &= \max \limits_{r=1,\dots,5}(\mathbf{w}_r y + \mathbf{b}).
    \end{aligned}
\end{equation}

\begin{figure}[h!tb]
    \subfigure[tanh]{
    \includegraphics[width=0.45\textwidth]{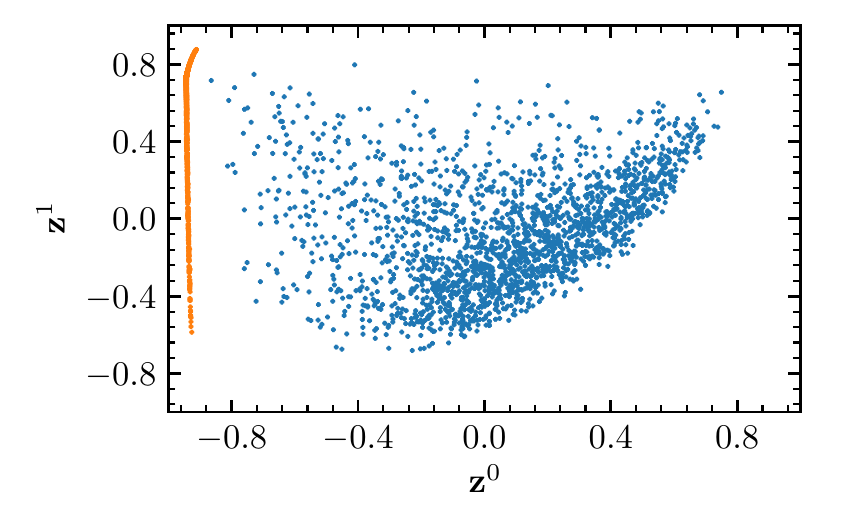}
    \label{fig:AE_code_tanh}}\hspace{-0em}
    ~
    \subfigure[sigmoid]{
    \includegraphics[width=0.45\textwidth]{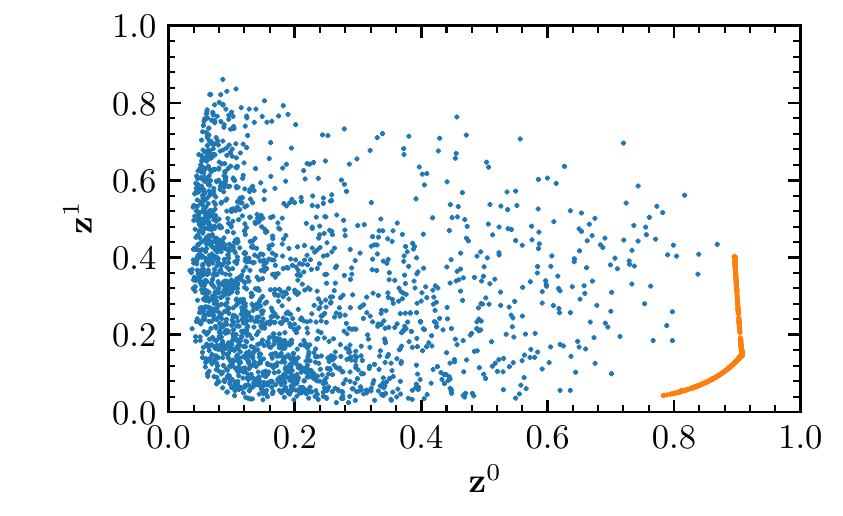}
    \label{fig:AE_code_sigmoid}}\hspace{-0em}
    ~
    \subfigure[maxout]{
    \includegraphics[width=0.45\textwidth]{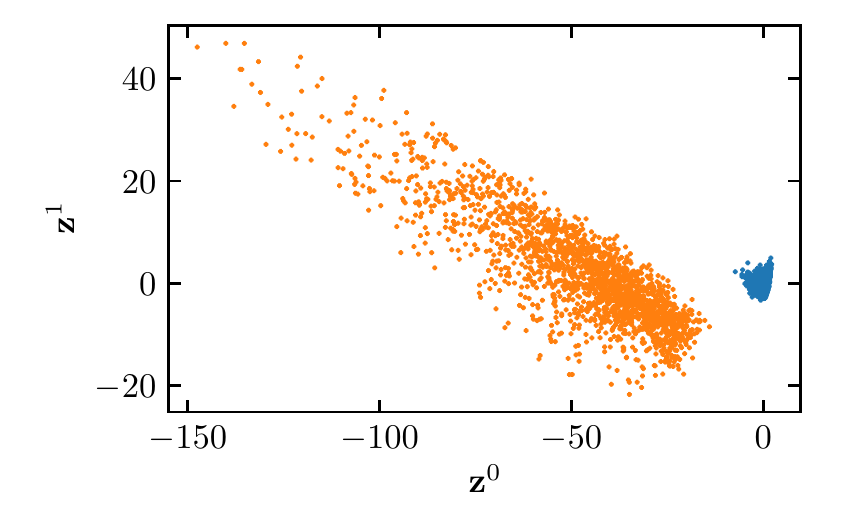}
    \label{fig:AE_code_maxout}}\hspace{-0em}
    ~
    \subfigure[ReLU]{
    \includegraphics[width=0.45\textwidth]{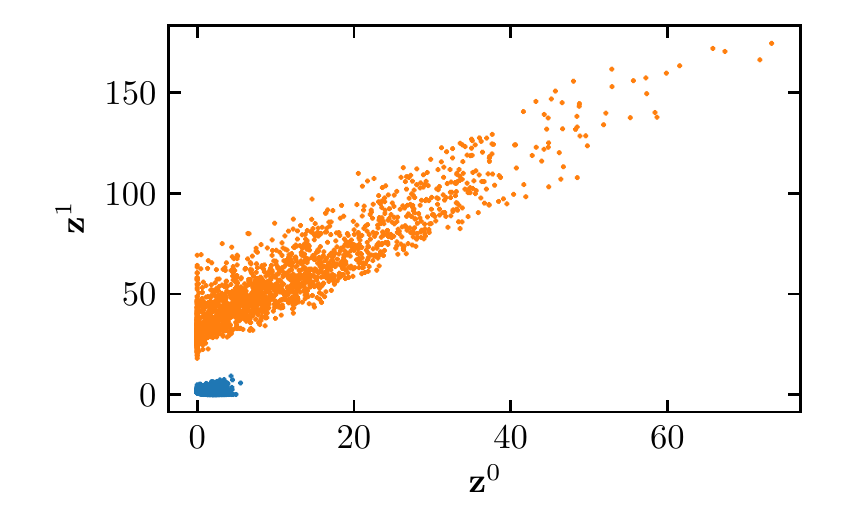}
    \label{fig:AE_code_relu}}\hspace{-0em}
    \caption{\emph{Good} (blue dots) and \emph{bad} data (orange dots) in 
             code space of AEs with $\mathcal{C} = \{12, 2, 12 \}$ and different
             activation functions.}
    \label{fig:AE_code}
\end{figure}

Figures \ref{fig:AE_code} shows $1000$ data points of the sets $\mathcal{X}^{g}$ and $\mathcal{X}^{b}$
each, mapped into the latent code space of AEs with these activation functions. The resulting sets 
$\mathcal{Z}^{g}$ and $\mathcal{Z}^{b}$ are colored in blue and orange respectively. Using tanh or 
sigmoid as activation functions, $\mathcal{Z}^{g}$ and 
$\mathcal{Z}^{b}$ appear difficult to separate. A large fraction of the \emph{good} data points are mapped 
into a ellipsoid-shaped cluster for the tanh activation function whereas using sigmoids maps them into a
hyperbola-shaped cluster. Data from $\mathcal{Z}^{g}$ however show significant scatter around their
respective clusters. \emph{Bad} data are mapped onto band-like structures at the boundary of the image
domain of the respective activation functions.
Using Maxout or ReLU activation functions, the AE maps \emph{good} data points into a narrow
cluster and scatters \emph{bad} data points along band-like structures.
The image domain of these activation functions has no upper bound such that the separation of
the \emph{good} and \emph{bad} data is larger for Maxout and ReLU activation functions than it
is for tanh and sigmoids.
The ReLU activation function can be considered as a special case of the Maxout function.
Since the results obtained by those two activation functions are qualitatively comparable and since
Maxout introduces additional parameters, in the following we only consider AEs using ReLU 
activation functions.

Codes produced by AEs with different layouts are qualitatively similar to those shown in \Figref{AE_code}.
For AEs with $z = 3$, the data points usually feature only little variance along one of
the three dimensions. That is, they cluster in a similar manner as they do for AEs with $z=2$.
Introducing an additional bottleneck layer in the AE, i.~e.~ choosing $\mathcal{C} = \{12, 5, 2, 5, 12\}$,
we observe a similar clustering of the data as is the case for $\mathcal{C} = \{12, 2, 12\}$.
Postponing the effect produced by different configurations of $\mathcal{C}$ on the resulting statistics of the inlier $\Te$ data, we continue by discussing the choice of a 
vector classifier in the code space $\mathcal{Z}$ of the AE.

Once an AE is trained, it defines a mapping from the input domain $\mathcal{X}$ into a unique, latent code space
$\mathcal{Z}$. A classifier is trained on $\mathcal{Z}^{g}$ and $\mathcal{Z}^{b}$ and subsequently used to assign
each $\mathbf{x} \in \mathcal{X}^{u}$ a label $\ell \in \{ \emph{good}, \emph{bad} \}$. The set of all
labels for the elements of $\mathcal{X}$ is denoted as $\mathcal{L}$. A label $\ell \in \mathcal{L}$ denotes whether a sample will be considered as an inlier
or outlier respectively. Such a classification introduces a bias, but with a validation procedure it is
possible to evaluate how well it generalizes to unseen data and select the most suitable model accordingly.

Here we consider three standard classifiers for vectorial data: a support vector machine
classifier (SVC), a nearest prototype classifier (PROT), and a so-called least-squares classifier (LS).
The details of these classifiers are provided in the appendix.

\begin{figure}[h!tb]
    \centering
    \includegraphics[keepaspectratio,width=0.5\textwidth]{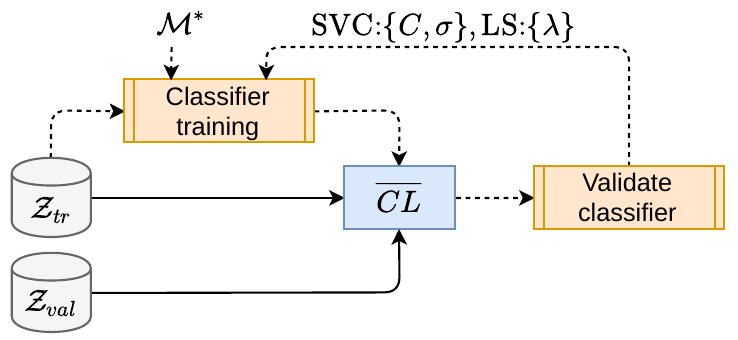}
    \caption{Pipeline of the classifier training. The classifier $\overline{CL}$ trained on $\mathcal{Z}_{tr}$ 
             is validated on $\mathcal{Z}_{val}$, to test its generalization capability and choosing the
             hyperparameters (such as $C$ and $\sigma$ in the SVM case). The model of the classifier 
             $\mathcal{M} \in \{ \text{SVC}, \text{LS}, \text{PROT} \}$ is assumed to be given at this stage. }
    \label{fig:class_train}
\end{figure}

To train a classifier, data is partitioned into a training and a validation set,
$\mathcal{Z}_\text{tr}$ and $\mathcal{Z}_\text{val}$. 
These sets contain only labelled samples: 
$\mathcal{Z}_\text{tr} = \{ \mathcal{Z}^{g}_\text{tr} \cup \mathcal{Z}^{b}_\text{tr} \}$, 
$\mathcal{Z}_\text{val} = \{ \mathcal{Z}^{g}_\text{val} \cup \mathcal{Z}^{b}_\text{val} \}$.
The \emph{good} training and validation data sets contain 1000 random data points each and the \emph{bad}
training and validation data sets contain approximately half the \emph{bad} data each. 
$\mathcal{Z}_\text{tr}$ is used to train the classifier, $\mathcal{Z}_\text{val}$ is used to evaluate the
generalization capability of the classifier. Fig.~\ref{fig:class_train} provides a schematic depiction of the
pipeline to train the classifier.

The generalization capability of the classifier is quantified by the so-called $F1$ score.
It is defined as the harmonic mean of precision and recall, as calculated for the validation data,
and assumes a value between zero and one. Precision is defined as the ratio of correctly classified
outliers and all correctly classified data points. Recall is defined as the ratio of correctly
classified outliers and the number of all data points classified as outliers. An $F1$ score of zero
describes a perfectly inaccurate classifier and a F1 score of one describes a perfectly accurate classifier.

\begin{figure}[h!tb]
    \centering
    \includegraphics{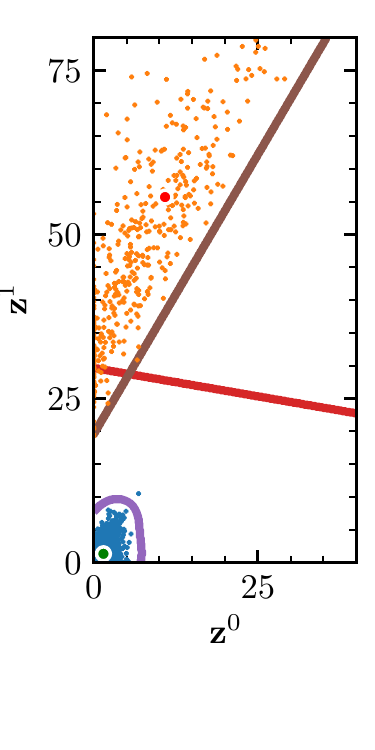}
    \caption{Decision boundaries for a nearest prototype classifier (red line), a support vector
             machine classifier (purple line), and a least-square classifier (brown line). The
             red and green circle denote the class prototypes given by \Eqnref{representatives}.
             The blue dots denote data from $\mathcal{Z}_\text{tr}^{g}$ and the orange dots denote
             data from $\mathcal{Z}_\text{tr}^{b}$. The green and red circle denote the prototypes
             given by \Eqnref{representatives}.}
    \label{fig:decision_boundary}
\end{figure}

Figure \ref{fig:decision_boundary} shows the decision boundaries learned by the three
different classifiers as full lines. The training data used to learn the decision boundaries
$\mathcal{Z}_\text{tr}$ are indicated by the blue and orange dots.
The SVC classifier, indicated by the purple line, draws a tight and curved decision boundary around $\mathcal{Z}^{g}_\text{tr}$, 
and the least-square classifier, the full brown line, draws a tight, linear boundary around $\mathcal{Z}^{b}_\text{tr}$.
The decision boundary identified by the nearest prototype classifier, the red line, puts the
decision boundary approximately half way between the class prototype. The F1-score of the
classifiers are respectively given by $1.0$, $1.0$ and $0.97$ for the shown data. This suggests that
all three classifiers correctly label unseen data as either \emph{good} or \emph{bad}, that is,
all three classifiers generalize equally well to unseen data.

\begin{figure}[h!tb]
    \centering
    \includegraphics{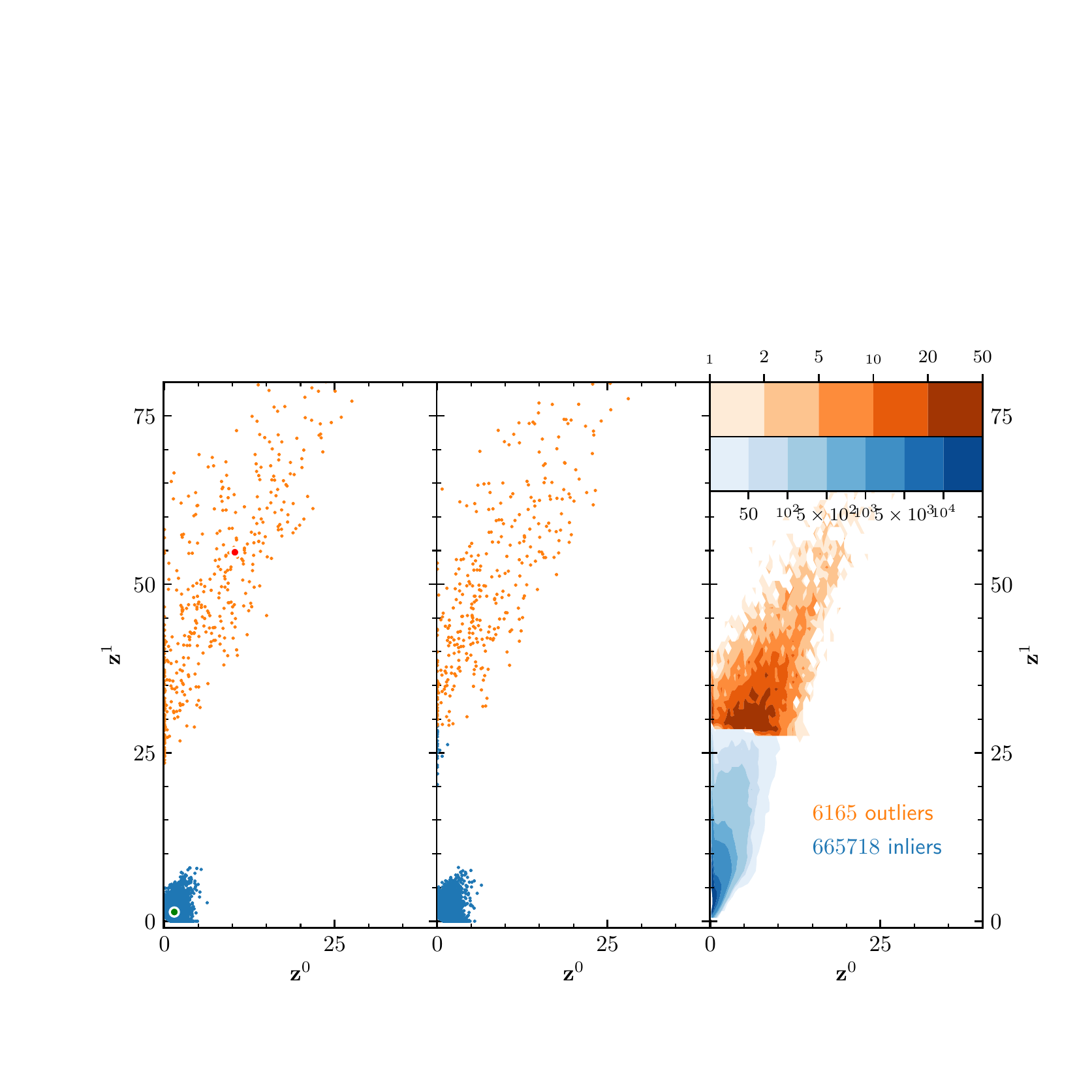}
    \caption{Representation of the MLP data in code space of an AE with
             a single hidden layer of size $z=2$.
             Blue denotes valid data, orange denotes invalid data points.
             The left and the middle panel respectively show training data $\mathcal{Z}_\text{tr}$
             and validation data $\mathcal{Z}_\text{val}$ for
             the classifier \Eqnref{representatives}. The right panel shows the count of data samples
             classified as either \emph{good} (blue) or \emph{bad} (orange).}
    \label{fig:code_clf}
\end{figure}

Figure \ref{fig:code_clf} shows an example of the classification process using the
nearest prototype classifier. The leftmost
panel shows the codes $\mathcal{Z}_\text{tr}^g$ in blue dots and the codes $\mathcal{Z}_\text{tr}^b$
in orange dots. The codes are clearly linearly separable, there is large leeway for placing the
decision boundary. A nearest prototype classifier is fitted on $\mathcal{Z}_\text{tr}$, the
prototypes $\mu_g$ and $\mu_b$, as defined in \Eqnref{representatives},
are depicted by a green and red dot respectively. This classifier is
subsequently used to assign class labels to the validation data $\mathcal{Z}_\text{val}^g$ and
$\mathcal{Z}_\text{val}^b$, shown in the same color coding in the middle panel. Only few codes
are mislabelled by the classifier, its F1 score is approximately one. The rightmost panel shows
the count of \emph{uncertain} data codes $\mathcal{Z}_\text{u}$ with assigned class labels.

\begin{figure}[h!tb]
\centering
    \subfigure[Average electron temperature using $\mathcal{X}^{'g}$, grouped by
                 hyperparameters of the AE and MLP.]
                {
                    \includegraphics[width=0.45\textwidth]{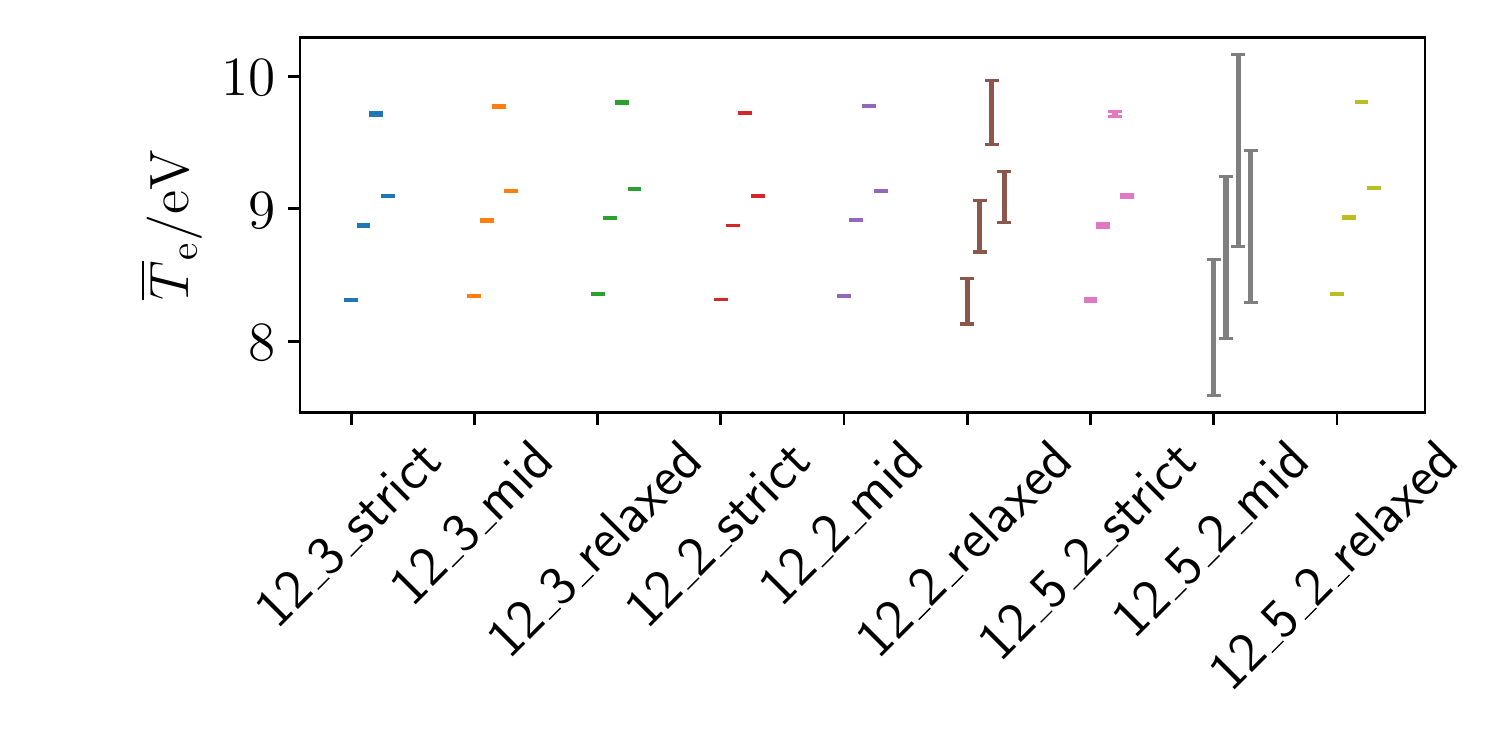}
                    \label{fig:training_te_inlier}
                }\hspace{-0em}
    ~
	\subfigure[Average relative fit error on the electron temperature using $\mathcal{X}^{'g}$, grouped by
                 hyperparameters of the AE and MLP.]
                {
                \includegraphics[width=0.45\textwidth]{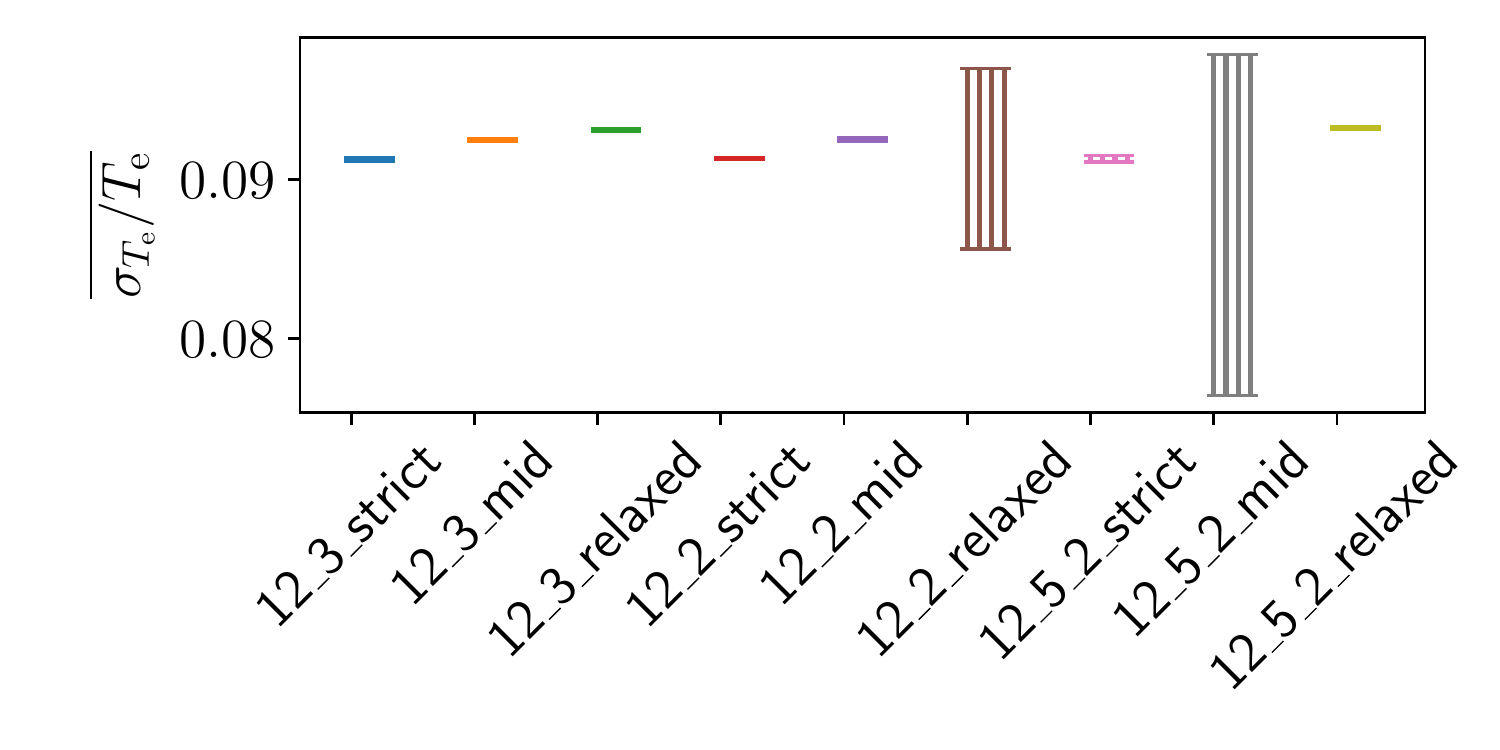}
                \label{fig:training_sigte_inlier}
                }
    \subfigure[Average electron temperature using $\mathcal{X}^{'b}$, grouped by
                 hyperparameters of the AE and MLP.]
                {
                \includegraphics[width=0.45\textwidth]{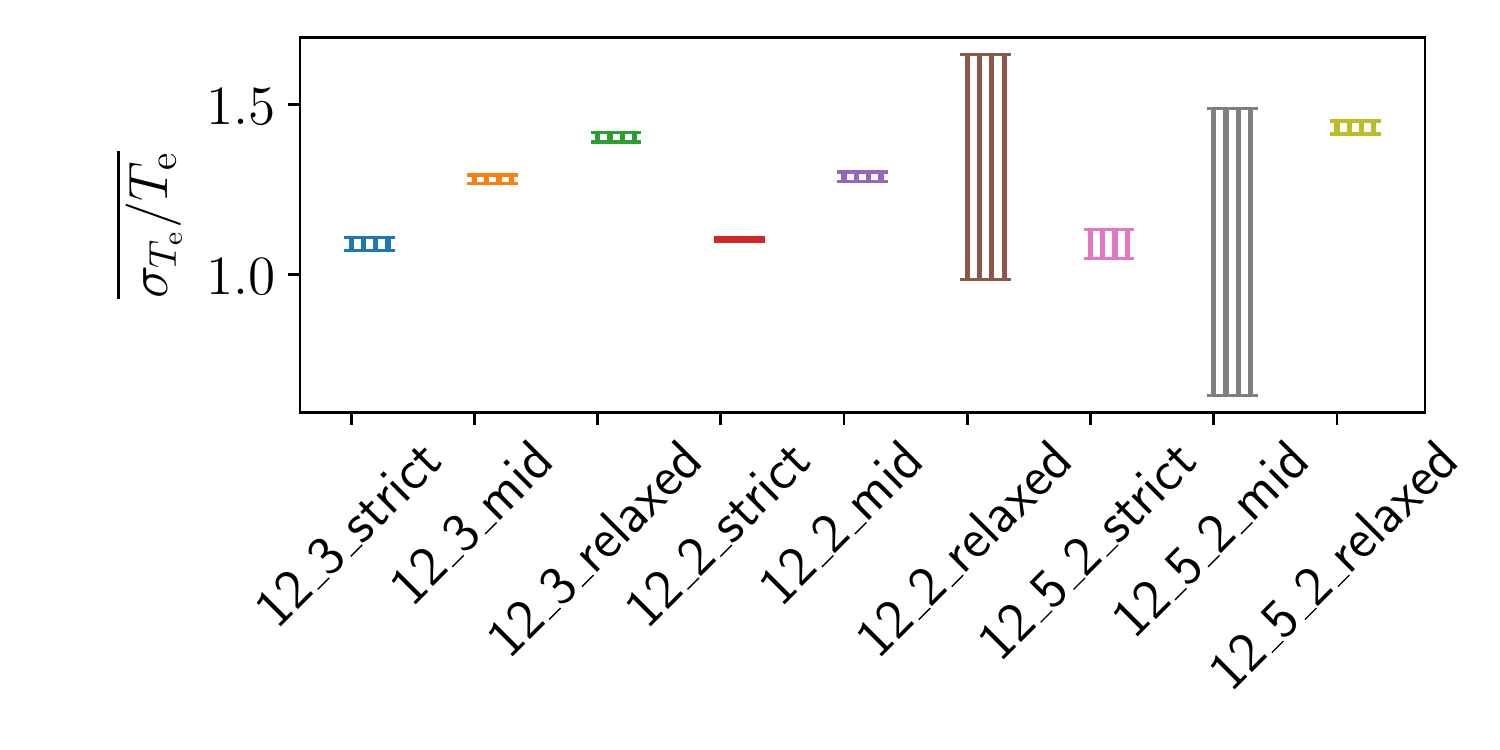}
                \label{fig:training_te_outlier}
                }\hspace{0em}
    ~
    \subfigure[Average relative fit error on the electron temperature using $\mathcal{X}^{'b}$, grouped by
                 hyperparameters of the AE and MLP.]
                {
                \includegraphics[width=0.45\textwidth]{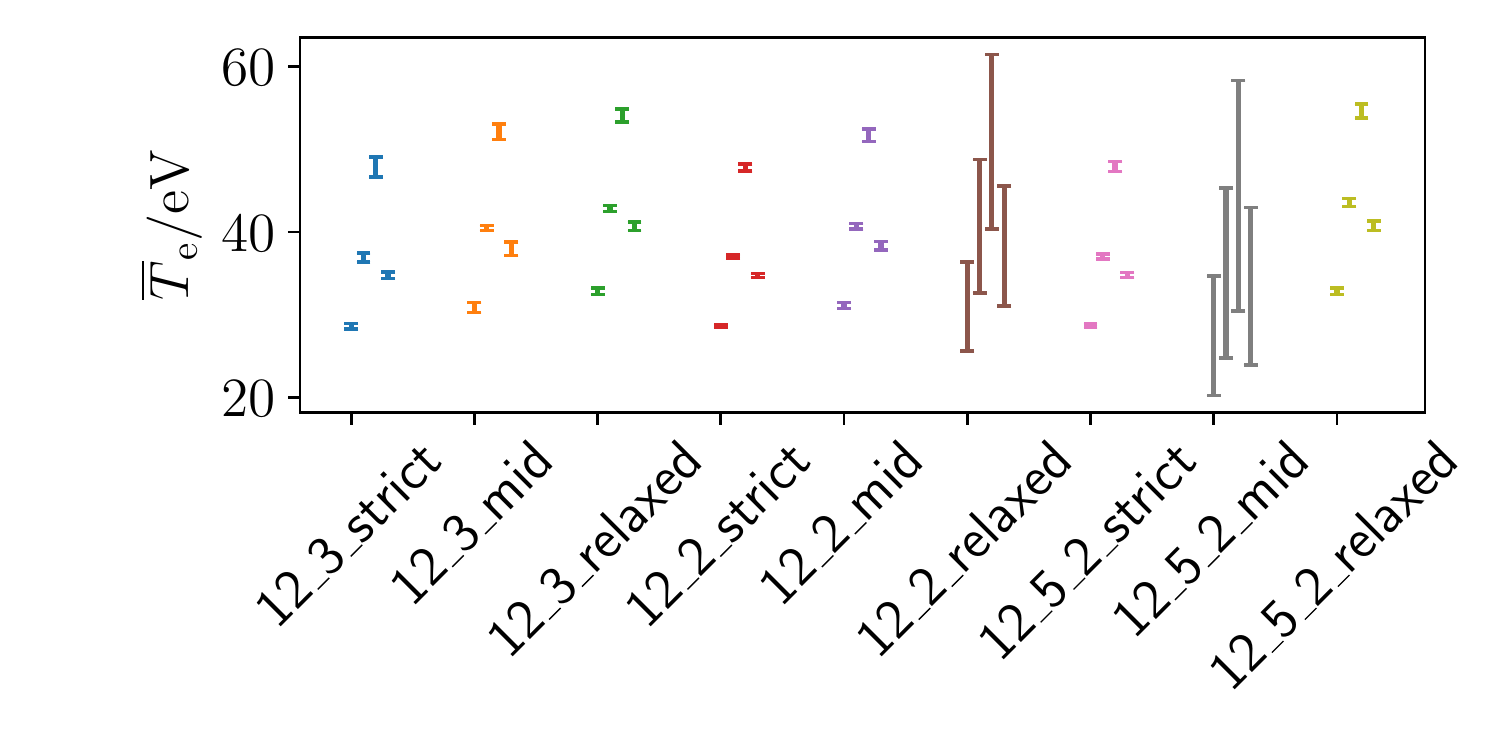}
                \label{fig:training_sigte_outlier}
                }
\caption{Average electron temperature and relative error on the electron temperature
         for the inlier and outlier samples, as identifed using different configurations
         of the AE and partition thresholds.}
\label{fig:results_ae_layout}
\end{figure}

Returning to the optimal configuration of the AE, we continue by discussing the statistics 
of all inlier samples
$\mathcal{X}^{'g} = \mathcal{X}^{g} \cup \{ \mathcal{X}^{u} | \mathcal{L}^{u} = \emph{good} \}$
and outlier samples
$\mathcal{X}^{'b} = \mathcal{X}^{b} \cup \{ \mathcal{X}^{u} | \mathcal{L}^{u} = \emph{bad} \}$,
as identifed by the proposed framework using the nearest prototype classifier.
Figures \ref{fig:training_te_inlier} - \ref{fig:training_sigte_outlier} show the average electron
temperature and the relative error on the electron temperature for different a-priori partitioning
and different AE layouts. The numerals in the x-axis labels denote the AE layout $\mathcal{C}$ and
staggered plot markers refer to data from the individual MLPs ``NE'', ``SE'', ``SW'', and ``NW''. 
The error  bars denote the sample standard variation.
For the inlier samples, $\overline{T}_\mathrm{e}$ varies between $8$ and $10\, \mathrm{eV}$.
This average shows little sensitivity to the used AE layout and the partition thresholds.
There also appears a systematic difference in $\Te$ as reported by the different probe heads.
This may be due to shadowing of plasma flows, caused by the protruding probe head geometry.
Plasma that is ballooned out at the outboard mid-plane will stream along the magnetic
field lines. Following the field lines, it impinges first on the west electrodes.
On the other hand, this discrepancy may also be due to a systematic error in the voltage
measurements among electrodes due to slightly untuned capacitor bridges in the electronics.

The root-mean-square values of the $\Te$ data are negligible for most $\mathcal{X}^{g}$, except for
the $\mathcal{C} = \{12, 2, 12\}$ layout using \textit{relaxed} partition thresholds
and the $\mathcal{C} = \{12, 5, 2, 5, 12\}$ layout using \textit{mid} partition thresholds.
This effect is due to randomness in the used input data for the AE training.
For these cases, significant root mean square values in $\mathcal{X}^{'g}$ are seen. 
Data points classified as outliers, $\mathcal{X}^{'b}$, show average electron temperatures
between approximately $30$ and $50\, \mathrm{eV}$. The relative error on these samples
is given by approximately one. Again, the standard deviation of these samples
is neglible in almost any AE configuration.
This analysis suggests that the choice of a specific AE layout does not yield significantly
different sample statistics of $\mathcal{X}^{'g}$. Therefore, we opt for the simplest
configuration $\mathcal{C} = \{12, 2, 12\}$.

\section{Performance evaluation}
\label{sec:performance}

\begin{figure}[h!tb]
    \centering
    \includegraphics[keepaspectratio,width=0.6\textwidth]{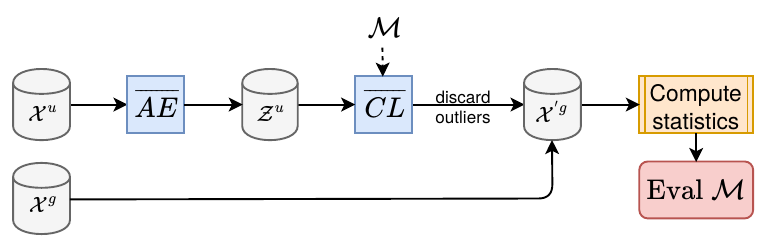}
    \caption{Uncertain data $\mathcal{X}_u$ are processed by $\overline{AE}$ and $\overline{CL}$, trained
             as described in the previous section. To evaluate the classification model $\mathcal{M}$,
             we compare statistics on the classification results $\mathcal{X}^{'g}$.}
    \label{fig:uncertain}
\end{figure}

In order to evaluate the performance of the proposed classification scheme, we continue by
comparing the statistics of the final inlier data set as well as the lower order statistical
moments of the heat flux, as computed from this data. The final inlier data sets are denoted by
$\mathcal{X}^{'g}$, and are respectively identified using
an SVC classifier, $\mathcal{X}^{'g}_\text{SVC}$, 
a least squares classifier, $\mathcal{X}^{'g}_\text{lsq}$,
and a nearest prototype classifer, $\mathcal{X}^{'g}_\text{pro}$. 
We evaluate their performance by comparing the resulting statistics to those obtained from
the entire dataset $\mathcal{X}$, the data without a-priori outliers, 
$\mathcal{X} \setminus \mathcal{X}^{b}$, as well as the data set of only a-priori inliners, $\mathcal{X}^{g}$.
\Figref{uncertain} illustrates
the processing pipeline used to obtain these data sets. For the results presented here, an AE with ReLU activation functions and 
$\mathcal{C} = \{ 12, 2, 12 \}$ is used.

%

\begin{figure}[h!tb]
\centering
    \subfigure[All data, $\mathcal{X}$]{
    \includegraphics[width=0.45\textwidth]{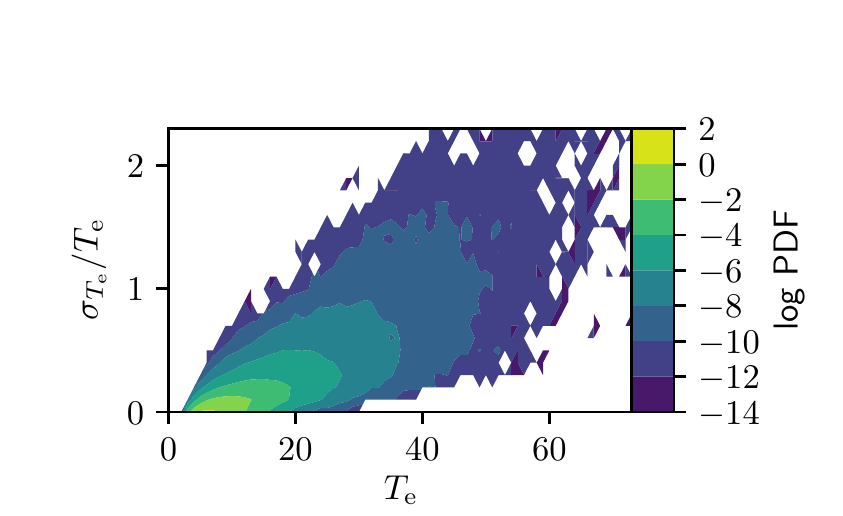}
    \label{fig:jointpdf_all}}\hspace{-0em}
    ~
	\subfigure[Only \emph{good} data, $\mathcal{X}^{g}$]{
    \includegraphics[width=0.45\textwidth]{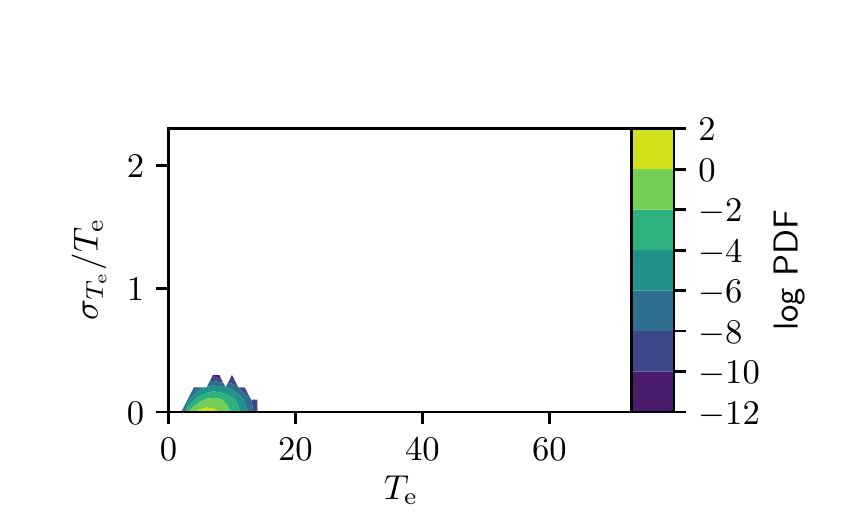}
    \label{fig:jointpdf_good}}
	~
    \subfigure[No \emph{bad} data, $\mathcal{X} \setminus \mathcal{X}^{b}$]{
    \includegraphics[width=0.45\textwidth]{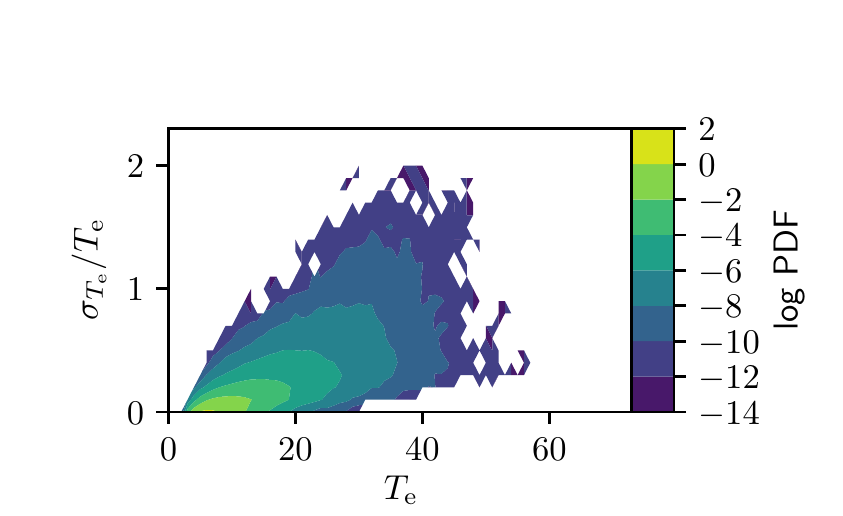}
    \label{fig:jointpdf_thr}}\hspace{0em}
    ~
    \subfigure[$\mathcal{X}^{'g}_\text{SVC}$]{
    \includegraphics[width=0.45\textwidth]{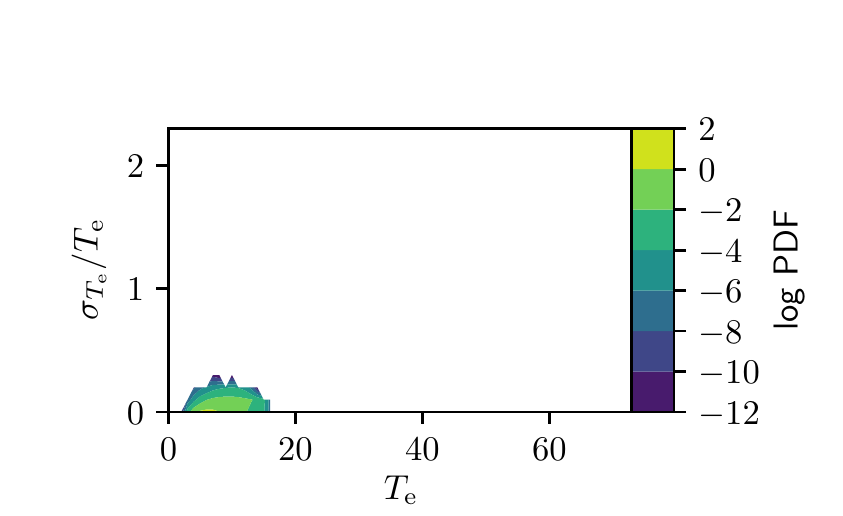}
    \label{fig:jointpdf_svc}}
    ~
    \subfigure[$\mathcal{X}^{'g}_\text{pro}$]{
    \includegraphics[width=0.45\textwidth]{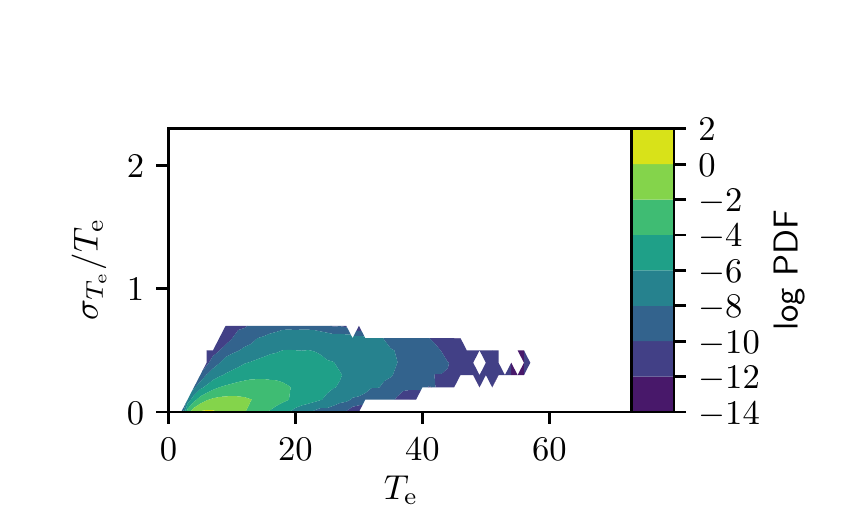}
    \label{fig:jointpdf_prt}}
    ~
    \subfigure[$\mathcal{X}^{'g}_\text{lsq}$]{
    \includegraphics[width=0.45\textwidth]{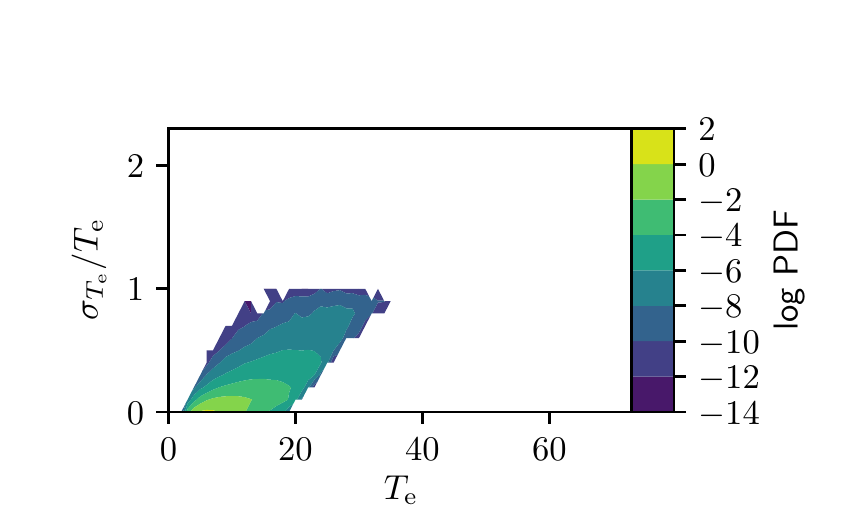}
    \label{fig:jointpdf_lsq}}
\caption{Joint probability distribution function of the average electron temperature and
         the average relative error on the electron temperature after outliers have been removed
         by different methods.}
\label{fig:jointpdfs}
\end{figure}

Figure \ref{fig:jointpdfs} shows the joint probability distribution function of the electron temperature
and the relative error on the electron temperature as computed for these data sets. Here, $\Te$ and 
$\sigte$ denote the average value reported by all four MLPs.
The entire data set $\mathcal{X}$, shown in \Figref{jointpdf_all}, features many samples with
small to medium $\Te$, associated with small to medium $\sigte$. A non-negligible fraction
of the samples however feature large $\Te$ values with $\sigte \gtrsim 1$. 
Considering only the good data, $\mathcal{X}^{g}$, shown in \Figref{jointpdf_good}, all samples
feature small $\Te$ values and a neglible relative error. The joint PDF of the set 
$\mathcal{X} \setminus \mathcal{X}^{b}$ is similar to that of the set $\mathcal{X}$,
but samples with $\Te \gtrsim 40\, \mathrm{eV}$ are almost absent. This is due to the
\emph{strict} threshold values applied when removing $\mathcal{X}^{b}$.

Pruning the MLP data using an SVC classifier, $\mathcal{X}^{'g}_\text{SVC}$, shown in \Figref{jointpdf_svc}, 
the joint PDF appears similar in shape to the one for $\mathcal{X}^{g}$, \Figref{jointpdf_good}. Only 
samples with $\Te \lesssim 15\, \mathrm{eV}$, associated with $\sigte \lesssim 0.3$ are
present. 
Removing outliers identified by the nearest prototype classifier, $\mathcal{X}^{'g}_\text{pro}$, shown
in \Figref{jointpdf_prt}, several samples with $\Te \gtrsim 50 \, \mathrm{eV}$ are present. However,
all samples feature relative errors less than approximately $0.75$. Qualitatively,
this joint PDF is similar to the joint PDF for $\mathcal{X} \setminus \mathcal{X}^{b}$,
\Figref{jointpdf_thr}, except that samples with large $\sigte$ are missing.
Employing a least squares classifier, $\mathcal{X}^{'g}_\text{lsq}$, shown in \Figref{jointpdf_lsq},
the resulting joint PDF is approximately aligned along an equi-probability contour of the joint PDF 
for $\mathcal{X}$, \Figref{jointpdf_all}. There are no samples with $\Te \gtrsim 35\, \mathrm{eV}$ and samples with 
$\sigte \gtrsim 1$ are also absent. Notably, samples $\Te \gtrsim 20 \mathrm{eV}$
with small $\sigte$ are absent while the data set still includes samples with
$\Te \gtrsim 20 \mathrm{eV}$ and large values of $\sigte$.

\begin{figure}[h!tb]
    \centering
    \includegraphics{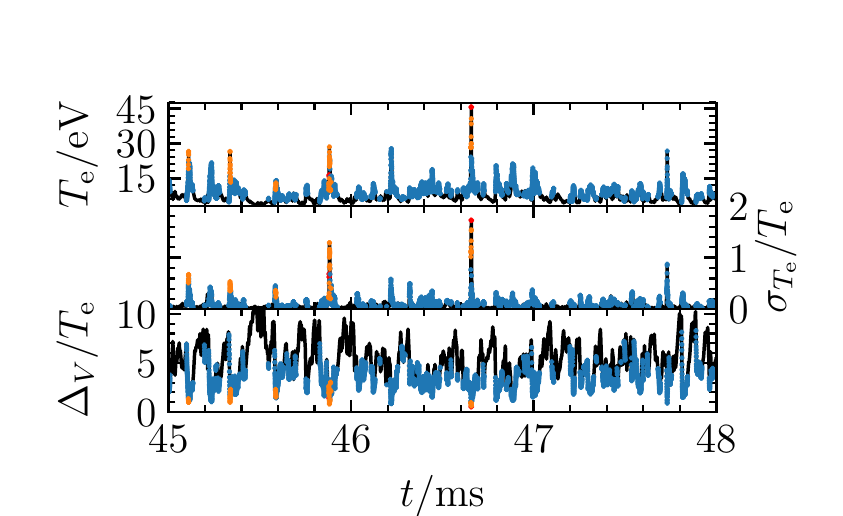}
    \caption{Data time series of the north-east MLP (cf.~ blue lines in \Figref{fit_timeseries}),
             overlaid with labels indicating classification of the data. Blue dots denote
             \emph{good} samples, $\mathcal{X}^{'g}_{\text{pro}}$, orange dots denote \emph{bad} samples,
             $\mathcal{X}^{'b}_{\text{pro}}$, and red crosses denote invalid samples $\mathcal{X}^{b}$, as
             classified by the prototype classifier using \emph{strict} thresholds.}
    \label{fig:timeseries_clf}
\end{figure}

Figure \ref{fig:timeseries_clf} shows the mapping of the labels $\mathcal{L}_\text{te}$, as
identified by the nearest prototype classifier into the time domain. The black lines and the 
red dots denote $\mathcal{X}^g$ and $\mathcal{X}^b$ respectively. Blue dots mark samples from
$\mathcal{Z}_\text{u}$
labelled $\ell = \textit{good}$, orange dots mark samples from $\mathcal{Z}_\text{u}$
labelled $\ell = \textit{bad}$. The large amplitude fluctuations, at $45.1\, \mathrm{ms}$, at
$45.9\, \mathrm{ms}$, and at $46.6\, \mathrm{ms}$ are mostly classified as \textit{good} data
points. Notably, the peak at $46.2\, \mathrm{ms}$ is classified as \textit{good}, even though
the relative error and the range of the biasing voltage of this MLP are similar to the conditions
of the preceding peak at $45.9\, \mathrm{ms}$. This is due to the requirement that at least two
MLPs need to report a invalid fit in order for a data point to be rejected.

A unique capability of Mirror Langmuir Probes is that they allow to study the fluctuation 
statistics of plasma flows driven by the electric drift. The heat flux impinging on plasma
facing components is of special interest. It is comprised of a conduction driven part,
$\widehat{\Gamma}_{T, \text{cond}} = \widetilde{U} \widetilde{T}_\mathrm{e} \langle n_\mathrm{e}  \rangle_\text{mv} / n_{\mathrm{e, mrms}}$,
a convection driven part
$\widehat{\Gamma}_{T, \text{conv}} = \widetilde{U} \widetilde{n}_\mathrm{e} \langle T_\mathrm{e}  \rangle_\text{mv} / T_{\mathrm{e, mrms}}$,
and contributions from triple correlations
$\widehat{\Gamma}_{T, \text{tcor}} = \widetilde{U} \widetilde{n}_\mathrm{e} \widetilde{T}_{\mathrm{e}}$.
Here $\widetilde{\cdot}$ denotes a quantity re-scaled by subtracting its moving average,
$\langle \cdot \rangle_{\textrm{mv}}$, and dividing by its moving root-mean-square $\cdot_\text{mrms}$.
In the following, we use a window length of $16384$ elements for these filters \cite{kube-2018-ppcf}.

\begin{figure}
    \centering
    \includegraphics{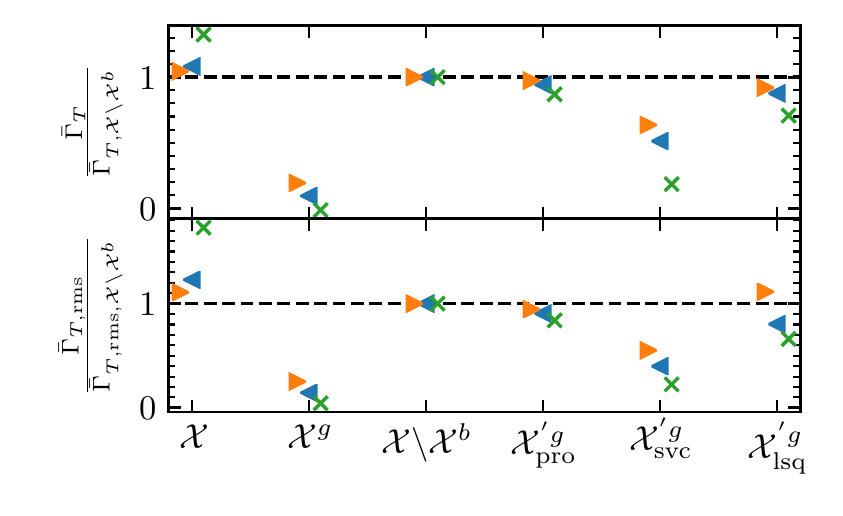}
    \caption{Radial heat flux due to conduction (triangle left), convection (triangle right), and triple correlations
             (cross), as computed from the various data sets and relative to reference values computed using
             $\mathcal{X} \setminus \mathcal{X}^b$. The upper panel shows the sample average, the lower panel
             shows the sample standard deviation.}
    \label{fig:clf_heatflux}
\end{figure}

Figure \ref{fig:clf_heatflux} shows the sample average and standard deviation for the
three contributions of the radial heat flux, computed using different data sets and relative
to the statistical moments computed ignoring a-priori outliers $\mathcal{X} \setminus \mathcal{X}^{b}$.
The average heat fluxes and the standard deviations are largest when using the entire dataset $\mathcal{X}$.
Using only good data, $\mathcal{X}^{g}$, yields averages and standard deviations less then $25\%$
of the values calculated using $\mathcal{X} \setminus \mathcal{X}^{b}$. Notably, for this data
the average radial heat flux due to triple correlations vanishes. Computing the moments using 
$\mathcal{X}^{'g}_\text{SVC}$, the average conductive and convective heat fluxes are approximately
$50$ and $60$ percent of the reference values, while the average value of the contributions from
triple correlations is approximately $20$ percent.

Removing outlier data as identified by the nearest prototype classifier, $\mathcal{X}^{'g}_\text{pro}$,
the average and root-mean-square values of the heat fluxes are approximately $85-95\%$ of the reference value.
Finally, using the least squares classifier results in statistical moments of the heat flux comparable to
those using the reference case $\mathcal{X} \setminus \mathcal{X}^{b}$.


The difference in the sample averages and standard deviations of the various
heat flux contributions can be related to the shape of the joint PDFs shown in \Figref{jointpdfs}.
For this, we note that the heat flux is computed from $\Te$, $\nee$ and $V_\mathrm{p}$ samples. The
relative error on $\nee$ is given by the geometric mean of the relative errors on $\Isat$ and $\Te$.
As discussed in \secref{measurements}, $\sigte$ and $\sigma_{\Isat} / \Isat$ are strongly correlated.
That is, a larger relative error on $\sigte$ implies a large relative error on the electron density.

Comparing the joint PDFs in \Figref{jointpdfs}, it is obvious that employing the different classifiers
to remove outliers, introduces slightly different biases into the inlier dataset.

%

%


\section{Conclusion}
\label{sec:conclusion}
In conclusion, we propose a framework to classify outlier data in data time series sampled
by a group of Mirror Langmuir probes in scrape-off layer plasmas. An autoencoder is trained
to identify a low-dimensional representation of \emph{good} fit data from this group of probes.
In this space, each dimension corresponds to a combination of features which best characterizes
the measurements. These are determined by the AE from the training data and without making any
a-priori assumption about the data set at hand. Outlier data, which does not share the characeristics
of \emph{good} data, appears in a separable cluster in the space of the AE. Several classifiers are
trained to separate outlier data in this space. With no ground truth available, the performance
of the classifiers are evaluated by comparing the lower order statistical moments of the radial
electron heat flux. 

Using either a least squares or a nearest prototype classifier results in similar 
statistics of the radial heat flux as obtained when using a threshold classifier to identify
outliers. Average contributions of the conductive and convective radial heat flux obtained by
these classifiers fall approximately $3$ and $14\%$ percent below the values obtained by applying
a threshold. On the other hand, the contribution due to triple correlations falls up to $40\%$
below the value obtained from the thresholding method. These differences result from the different
characteristics of the data points which are identified as outliers. While the least squares
classifier places the decision boundary close to the outlier data cluster, the nearest prototype
classifier places the decision boundary approximately equidistant to both clusters. That is,
the least squares classifier gives a more relaxed outlier removal while the nearest prototype
classifier has a lower threshold. 
While neither method can be identified as the correct method to remove outliers from the data set,
this study implies that not employing outlier removal may lead to heat fluxes, over-estimated
by a significant amount.

The framework proposed here may also be adapted to other types of sensors than MLPs. The requirements 
for applying the method describes here are first, that any single sensor reports a physical quantity
together with an uncertainty of that measurement. And second, any sensor in the group needs to sample
roughly the same environment.

\section*{Acknowledgements}
This work was supported with financial subvention from the Research Council of Norway under Grant No.
240510/F20 and the U.S. Department of Energy, Office of Science, Office of Fusion Energy Sciences,
using User Facility Alcator C-Mod, under Award No. DE-FC02-99ER54512-CMOD. 
F.~M.~B. is founded by the Research Council of Norway under FRIPRO Grant No. 239844 ``\emph{Next Generation Learning Machines}''.
R.~K.~ acknowledges the generous hospitality of the MIT Plasma Science and Fusion Center where parts of this
work were conducted.

\section*{Appendix}
Classifiers are algorithms that assign a class label $\ell$ to new data, on the basis of
previously seen training data with known class labels. For the case of two distinct classes,
we assign $\ell = \pm 1$ to inliers and outliers respectively. In the following, we describe how labels
$\ell_i$ for new data $\{ x_i \}, x_i \in \mathbb{R}^{d}$ are retrieved using the classifiers used
in this paper. 

\subsection*{Support Vector Machine~\cite{vapnik1999overview}}

A Support Vector Machines (SVM) learns a linear classifier in a kernel space,
\begin{equation}
\ell_i = g(x_i) = \text{sign}(\phi(w) \cdot \phi(x_i) + b),
\end{equation}
induced by a usually non-linear kernel $\phi(w) \cdot \phi(x) = K(w,x)$.
A typical choice for $K$ is the radial basis function, defined as 
\begin{equation*}
    K(x_i, x_j) = \exp \left( - \frac{\| x_i - x_j \|^2}{2 \sigma^2} \right).
\end{equation*}
where $\sigma$ is called the bandwidth of the kernel.

In order to train a SVM, the cost function
\begin{align*}
\phi^*(w) = \arg \min_{\phi(w)} \frac{1}{2} \left\lVert \phi(w) \right\rVert^2
\end{align*}
is minimized under the constrain that $y_i (\phi(w) \cdot \phi(x_i) + b) \ge 1$.
That is, the training data is taken to be only inlier data with $\ell_i = +1$.

%


The constraints can be included in the previous quadratic cost by using the Lagrangian multipliers, 
\begin{equation}
\label{eq:lagrangian}
L(\phi(w),b,\alpha) = \frac{1}{2}||\phi(w)||^2 - \sum_i \alpha_i (y_i (\phi(w) \cdot \phi(x_i) + b) - 1).
\end{equation}

It follows that the weight vectors become a linear combination of the data points
\begin{equation}
\phi(w) = \sum_i y_i \alpha_i \phi(x_i),
\end{equation}
and the classifier can be expressed as
\begin{equation}
\label{eq:new_class}
g(x) = \text{sign}\bigg(\sum_i y_i \alpha_i \phi(x_i) \cdot \phi(x) + b\bigg) = \text{sign}\bigg(\sum_i y_i \alpha_i K(x_i,x) + b\bigg).
\end{equation}

If we substitute \eqref{eq:new_class} in \eqref{eq:lagrangian}, we obtain the following dual cost function
\begin{equation}
W(\alpha) = \sum_i \alpha_i - \frac{1}{2} \sum_{i,j} y_i y_j \alpha_i \alpha_j \phi(x_i) \cdot \phi(x_j) = \sum_i \alpha_i - \frac{1}{2} \sum_{i,j} y_i y_j \alpha_i \alpha_j K(x_i,x_j),
\end{equation}
and the optimization now reads
\begin{equation}
\begin{aligned}
\hat{\alpha} = \arg \max_\alpha W(\alpha), \\
\text{such that} \quad \alpha_i \ge 0.
\end{aligned}
\end{equation}

Once the training is complete, new points are classified directly by applying \eqref{eq:new_class}.

\subsection*{Prototype classifier~\cite{bezdek2001nearest}}

Classification by means of simple a nearest prototype classifier, operates as follows.
For each class $c$, a prototype is computed as
\begin{equation}
    \mu_c = \frac{1}{|\{ x_i \}|} \sum \limits_{i} x_i.
    \label{eq:representatives}
\end{equation}
The class label $\ell$ of an uncategorized data sample $z$ is assigned as
\begin{align}
    \ell = \argmin_{c} \| z - \mu_c \|^2 \label{eq:classifier}
\end{align}

This classifier does not depend on any hyperparameter and requires to maintain only the representative
of each cluster to classify out-of-sample data. Due to its simplicity, this classifier cannot identify
complex decision boundaries to separate samples of different classes.

\subsection*{Least square classifier}

A classification function $f$ is learned by minimizing the following quadratic cost
\begin{equation}
\label{eq:ls_loss}
    \min \limits \frac{1}{N} \sum \limits_{i=1}^N \lVert \ell_i - f(x_i) \rVert^2 + \lambda \lVert f \rVert^2
\end{equation}
where the first term penalize the discrepancy between the output class of the function and the known
class of the training data. The second cost term instead encourage smoothness in the target function
and is useful to prevent overfitting.

In the analysis presented here, we choose $f$ to be a linear function
\begin{equation*}
    f(x) = Wx + b
\end{equation*}
whose parameters $W$ and $b$ are optimized according to \eqref{eq:ls_loss}.

A more flexible choice consists in using a kernel function to define $f$:
\begin{equation*}
    f(x) = \sum_{i=1}^N c_i K(x, x_i)
\end{equation*}

In this case, the objective of the optimization is to find the parameters $c_i$. This is done by modifying the quadratic loss in \eqref{eq:ls_loss}, which becomes
\begin{equation}
    \min \limits_{\mathbf{c}} \frac{1}{N} \sum \limits_{i=1}^N \lVert y_i - \sum_{j=1}^N c_j K(x, x_j) \rVert^2 + \lambda \lVert \mathbf{c}^T\mathbf{K}\mathbf{c} \rVert^2.
\end{equation}

\bibliography{myrefs}

\end{document}